\documentclass[%
 aip,
 amsmath,amssymb,
 reprint,%
]{revtex4-1}

\usepackage{gensymb}
\usepackage{graphicx}
\usepackage{dcolumn}
\usepackage{bm}
\usepackage{multirow}
\usepackage{color}
\usepackage{siunitx}
\usepackage{amsmath}

\usepackage[utf8]{inputenc}
\usepackage[T1]{fontenc}
\usepackage{mathptmx}

\begin{document}

\preprint{AIP/123-QED}

\title{Evidence of polyamorphic transitions during densified $\mathrm{SiO_{2}}$ glass annealing}

\author{Antoine Cornet}

 \altaffiliation{Now at the University of Manchester: antoine.cornet@manchester.ac.uk}
\affiliation{ Institut Lumi\`{e}re Mati\`{e}re, Univ Lyon, Universit\'{e} Claude Bernard Lyon 1, CNRS, F-69622 Villeurbanne, France
}%
\author{Christine Martinet}%

\affiliation{ Institut Lumi\`{e}re Mati\`{e}re, Univ Lyon, Universit\'{e} Claude Bernard Lyon 1, CNRS, F-69622 Villeurbanne, France
}%
\author{Val\'{e}rie Martinez}%

\affiliation{ Institut Lumi\`{e}re Mati\`{e}re, Univ Lyon, Universit\'{e} Claude Bernard Lyon 1, CNRS, F-69622 Villeurbanne, France
}%

\author{Dominique de Ligny}

\affiliation{
Department  of Materials Science, Glass and Ceramics, University Erlangen-N\"{u}rnberg, Martensstra., D-91058 Erlangen, Germany
}%

\date{\today}

\begin{abstract}
In-situ X-ray scattering monitoring is carried out during temperature annealing on different densified $\mathrm{SiO_{2}}$ glasses. Density fluctuations and intermediate range coherence from x-ray scattering (SAXS) and diffraction (WAXS) evidence a maximum in their evolution at the same relaxation time. These extrema confirm the existence of an intermediate transitory disordered state between the two more ordered high and low density amorphous states. We propose that the existence of this transitory state confirms the existence of two mega basin in the energy landscape and therefore an amorphous-amorphous transition. Including older Raman results, we show that this intermediate disorder state implies similar mechanisms at all length scales from a few angstroms to 5 nm. 
\end{abstract}

\maketitle


\section{\label{sec:level1}Introduction}

Polyamorphism describes the possibility for a liquid to exist in two distinct amorphous phases, with possible transitions among these. It was first presented as an explanation for the negative melting slope $\mathrm{\frac{dT_{_{melting}}}{dP}}$ of some liquidus\cite{Rapoport1967}, in particular water. This concept has been extended to the glassy state of water, leading to the progressive discovery of three distinct amorphous structures for ice\cite{Mishima1985, Loerting2011}. The concept of polyamorphism became stronger when evidenced in other glasses, such as in the YAG-$\mathrm{Al_{2}O_{3}}$ system, or in amorphous Si and Ge\cite{Brazhkin2003, McMillan2004, Wilding2006, Machon2014, Tetsuya2004}. In the canonical network glass, silica, polyamorphism was used to justify the increase in the coordination number of silicon at high pressure \cite{Scandolo2007, Yarger2010, Koziatek2015, Prescher2017, Murakami2019}. In this process the range of pressure where this coordination change appears is very wide, in contrast to the sharp transition observed in water. The transformation from $\mathrm{SiO_{4}}$ to $\mathrm{SiO_{6}}$ units spreads over 20 GPa at least, depending on studies \cite{Sato2008, Sato2010, Salmon2015}, preventing direct assessment of polyamorphism. On the other hand, polyamorphism can be linked to the densification of silica glass, since this latter phenomenon can be addressed as a coexistence of pristine glass and fully densified high density amorphous glass \cite{Sonneville2012, Sonneville2013}. In this last case, the study of the structure of the densified glass is not self sufficient, because the mechanisms of the amorphous-amorphous transitions (AAT) and the origin of polyamorphism remain unsolved.\\

To overcome this ambiguity, we analyzed the structural transformations of densified silica glass during annealings at high temperature and atmospheric pressure.

The idea of studying the reverse transformation rather than densification appeared as early as 1963 in MacKenzie's work. \cite{MacKenzie1963} As pionering the work of MacKenzie was, the multiplicity of the densification conditions, including shear, as well as the absence of structural information, didn't allow him to evidence any identifiable specific transformation. Ardnt \textit{et. al.} took a closer look to the structure when relaxing isothermally silica glass samples that were previously densified at high pressure and high temperature\cite{Hummel1989, Arndt1991}. In these papers, the authors noted that the density and the refractive index of the Low density Amorphous (LDA) pristine state were recovered after annealings at temperatures lower than the glass transition temperature, e.g. 800\degree C compared to 1200\degree C. Moreover, when comparing the evolution of the refractive index and the density, they evidenced a departure from the expected linear relation, that they interpreted as a transitory state. Later, Surovtsev \textit{et. al.} performed 1h annealings on densified silica at different temperatures, allowing them to obtain a very broad range of densities\cite{Surovtsev2006}. A careful examination of the low frequency part of the Raman spectrum of the recovered samples allowed them to point out the existence of a sharp change in the relaxationnal part of the spectrum. They interpreted this change as a transition in the stiffness of the material. Because this transition appears at a density of \SI{2.26}{\g\per\cubic\centi\metre}, close to those of trydmite and cristobalite, they concluded that the transition arises from cluster crystal-like structured heterogeneities. In a previous paper\cite{Martinet2015}, we demonstrated that the density cannot be used as a reference of state parameter to study the structure of vitreous silica (and glasses in general), since different isochemical amorphous structures can have equal densities. Nevertheless, there are several hints that indicate that a polyamorphic transition might be visible during the annealing of densified silica glass. Lately, Guerette \textit{et. al.} evidenced a transition in the mechanical properties of silica during annealings of densified samples, visible through the evolution of the elastic moduli\cite{Guerette2018}. In this study, we aim to track the underlying structural changes at the possible amorphous-amorphous transitions during the relaxation at high temperature of densified samples, at the different length scales of the glass structure.\newline

In a recent paper\cite{Cornet2017}, we reported annealing experiments on densified silica samples prepared with different P-T paths:
\begin{itemize}
    \item The activation energy of the process, $\mathrm{255 \pm }$\SI{45}{\kilo\joule\per\mole}, is significantly lower than the activation energy of viscous flow at the same temperatures\cite{Doremus2002}, i.e. \SI{720}{\kilo\joule\per\mole}. This indicates different mechanisms at the structural level.
    \item The evolution of the density during the transformation (Supporting Information) is a monotonous decrease for all the samples tested, and so for all the compression parameters tested. This is consistent with previous measurements\cite{MacKenzie1963, Arndt1969, Arndt1991, Surovtsev2006, Guerette2018}, and indicates a global monotonous volume expansion of the glassy network.
    \item The transformation goes through an transitory activated state, characterized by an increase in the disorder within the rings structure of the IRO. An increase in the intensity and area of the so-called Raman D2 band is witnessed for all the samples tested: for different compression temperatures at an identical pressure of 5 GPa, and for different nominal pressures at room temperature. This band is commonly assigned to three membered rings in silica, therefore the increase of the D2 band in the Raman spectrum indicates a creation of such three membered rings. These small rings are characterized by a mean Si-O-Si angle of $\mathrm{129 \pm 1 \degree}$ (see Ref \onlinecite{Martinet2015} for  densified silica structure), which is at the small bound of the overall Si-O-Si angle distribution in densified silica glass\cite{Trease2017}, making them the most compact structure that can be found in the glassy network. The creation of compact structure in an overall expanding network is counter-intuitive, unless it implies an increase of the level of heterogeneity and disorder. For instance, in a simplified case, two 6 membered rings could transform to a small 3 membered one and a large 9 membered one. The creation of three membered rings is then associated with the creation of larger voids in the structure, where the expansion can take place.
\end{itemize}
    
More generally speaking, the transformation can be related to an increase in the inhomogeneity of the structure during the relaxation process. To characterize this transitory state more thoroughly, we performed X-Ray scattering to track the homogeneity of the glass structure up to several nanometers.

In Wide Angle X-Ray Scattering (WAXS), the First Sharp Diffraction Peak (FSDP) is a feature present in the diffractograms of glasses when compared to the diffractogram of there crystalline counterparts, and whom attribution has been debated \cite{Elliott1991a, Zaug2008, Crupi2015}. The models of Mei \textit{et al.} and Crupi \textit{et al.} suggest that the FSDP arises from the boundary between interstitial voids, corresponding to the rings of tetrahedron in silica \cite{Mei2008, Crupi2015}. In any case, the position of the FSDP indicates the characteristic distance of the intermediate range order (IRO), when its width tells us about the distance over which the IRO remains coherent \cite{Elliott1991a}. Thus, on this length scale, variations of the homogeneity should impact the width of the FSDP.

At a larger scale, over 1 to 5 nm, the electronic density fluctuations $\langle \Delta \rho \rangle$ are directly proportional to the intensity of the forward diffusion of X-Ray (of a volume V), i.e. with a zero scattering vector q \cite{Guinier1955, Levelut1967}.
\begin{equation}
	I\left( q=0 \right) \propto V \langle \left(  \Delta \rho \right) ^{2} \rangle
	\label{eq:SAXS_density}
\end{equation}

\begin{equation}
I\left( q=0 \right) = \underbrace{I\left( q=0,T=0 \right)}_{static} + \underbrace{A \times k_{b} T \times \left\langle  \dfrac{1}{\rho V^{2} _{l,\infty}} \right\rangle}_{I(q=0)_{phonons}, dynamic}
\label{eq:SAXS_density_bis}
\end{equation}

In a glass, it was demonstrated that a static term is introduced to take into account the frozen-in heterogeneities \cite{Levelut2005}. The static term represents the largest structural order in glasses, i.e. the spatial heterogenities in the elastic moduli at the origin of the Ioffe-Regel criterion \cite{Mizuno2014}.  A dynamic term accounts for the spatial fluctuations caused by the propagating phonons, and is thus temperature $T$ dependent ($A$ being a proportionality factor, $k_{b}$ the Boltzmann constant). This contribution $\mathrm{I(q=0)_{phonons}}$ is linked to the (spatially averaged $\left\langle \right\rangle$) density $\rho$ and longitudinal sound wave velocity (probed at infinite frequency) $V_{l,\infty}$.\newline
It has been demonstrated that the glasses recovered from compression at high temperatures possess structures that are otherwise inaccessible through thermal history only (or pressure history only) \cite{Svenson2017}. Thus further the static term will be taken as constant. However, we will use the extrapolation at the origin of the intensity of the diffractograms $I\left( q=0 \right)$ to extract the density fluctuations at large length scale, over 1-5 nm.

\section{\label{sec:level2}Experimental}

\subsection{\label{sec:level2_1}Densification at high pressure-high temperature}

To study the possible AAT, we performed compressions on three glassy samples from suprasil 300 commercial (OH free) silica. Following the procedure described in Ref. \onlinecite{Cornet2017}, three macroscopic cylinders were compressed in a Belt press at 5 GPa and heated simultaneously at 425\degree C, 750\degree C and 1020\degree C. During the compression, the temperature increase, around 1\degree C/s, starts only after the nominal pressure is reached. The temperature is then maintained for 10 minutes, and finally the sample is quenched by turning off the electrical current in the furnace. After three to five minutes, the pressure is released and the sample is recovered. From experience, we estimate that the sample temperature is less than 100 \degree C when the pressure starts to decrease. The homogeneity of each sample was checked using Raman mapping. No differences were detectable in the collected spectra between the centre and the edge of the samples.

The final density is measured by the buoyancy method in toluene, and the densification ratio obtained for the recovered samples compared to the pristine glass are respectively 9.2\% , 15.5\% and 16.5\%. Samples are thus labelled as Belt 9.2\%, Belt 15.5\% and Belt 16.5\%. The compression parameters with the recovered densities are listed in table \ref{table:compression}. A description of their structure is available in Ref. \onlinecite{Martinet2015}.\\
\begin{table}[h]
\begin{center}
\renewcommand{\arraystretch}{1.2}
\begin{tabular}{ccccc}
\hline
\hline
Pressure & Temperature & Density & Densification & \multirow{2}*{Label} \\
(GPa) & (\degree C) & ($\mathrm{g.cm^{-3}}$) & ratio $\frac{\rho - \rho_{0}}{\rho_{0}}$&  \\
\hline
5 & 425 & $2.40 \pm 0,03$ & 9.2\% & Belt 9.2\% \\
5 & 750 & $2.54 \pm 0,03$ & 15.5\% & Belt 15.5\% \\
5 & 1020 & $2.56 \pm 0,02$ & 16.5\% & Belt 16.5\% \\
\hline
\hline
\end{tabular}
\end{center}
\caption{compression parameters, recovered densities and denomination of the different samples. The densities were estimated via buoyancy measurements in toluene at room temperature.}
\label{table:compression}
\end{table}

\subsection{\label{sec:level2_2}X-ray scattering at high temperature}

Several combined in-situ SAXS and WAXS experiments were performed at the BM26 beamline in the ESRF in Grenoble (France), using a monochromatic beam operating at 12 KeV (\SI{1.033}{\angstrom}). The sample was heated in air using a micro-tomo furnace encaspulated by a aluminum lid with kapton or mica windows. The SAXS signal was recorded using a pilatus 1M detector placed 1.4 meter away from the sample, with the transmitted x-rays traveling through a primary vacuum, the recorded SAXS signal ranges from \SI{0.15}{\per\nano\metre} to \SI{7}{\per\nano\metre}. To work on the FSDP solely, we needed to be able to substract the rest of the structure factor. Thus, the WAXS signal was recorded with a pilatus 300K detector placed as close as possible to the sample to get an significant part of this structure factor: 8 to \SI{50}{\per\nano\metre}.

We performed different annealings between 750\degree C and 900 \degree C. The temperature was first increased  at 10\degree C/min up to the desired value, and then maintained for various durations ranging from one to several hours. SAXS and WAXS data were recorded during the heating and the isothermal annealing. The annealing temperatures were chosen to lie in the same range that these used for the Raman based experiments. A run was conducted without the sample to obtain the background signal, that originates from the window and air scattering, over the entire experiment. With the intensity of the beam determined before and after the furnace, the diffractogramms were normalized following Levelut's method \cite{Levelut2005}. It is then possible to extract the position at maximum intensity and the width of the FSDP from the WAXS spectra, and the intensity extrapolated at $\mathrm{q = 0 nm^{-1}}$ from the SAXS data. The full procedure to extract the data from the raw signal is detailed in the Supporting Information (hereafter noted SI).

\section{\label{sec:level3}Results}

\begin{figure}[h!]
\center
\includegraphics[width=9cm]{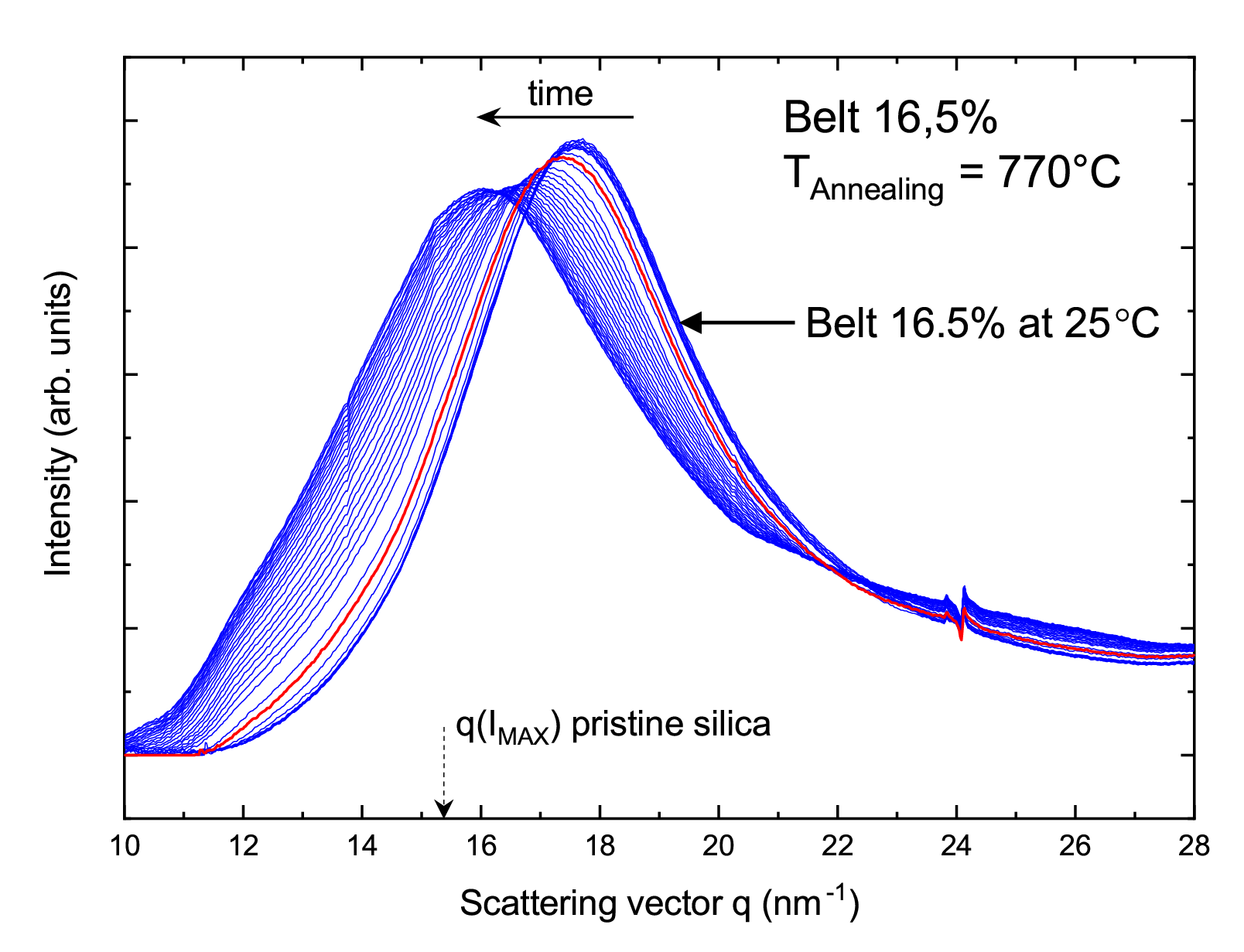}
\caption{Evolution of the FSDP for a Belt 16.5\% sample annealed at 770\degree C. The FSDP before annealing, at room temperature, is the one as the highest q at maximum intensity. The red FSDP corresponds to the first diffractogram recorded at the 770\degree C annealing temperature.}
\label{fig:FSDP_example}
\end{figure}

\begin{figure}[h!]
\center
\includegraphics[width=9cm]{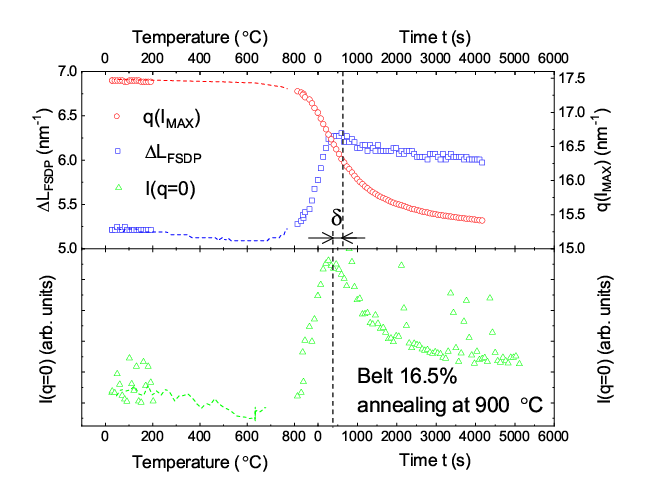}
\caption{Evolution of the position of the maximum $\mathrm{q \left( I_{MAX} \right)}$, of the half width $\mathrm{\Delta L_{FSDP}}$ of the FSDP and the intensity extrapolated at $\mathrm{q \ = \ 0 \ nm^{-1}}$ $\mathrm{I \left( q = 0 \right) }$ for the sample Belt 16.5\% annealed at 900 \degree C. The different parameters are expressed as a function of the temperature during the heating part, and as a function of the time during the isothermal annealing. The blank from 200\degree C to 800\degree C is due to a loss of the beam during the experimental session.}
\label{fig:DRX_Belt_16}
\end{figure}

\begin{figure}
\center
\includegraphics[width=9cm]{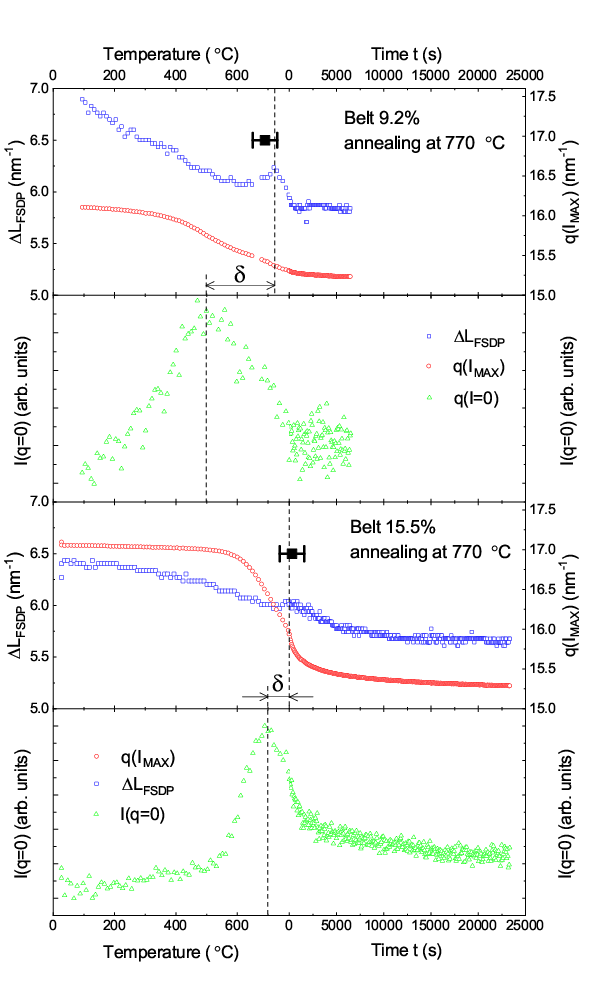}
\caption{Evolution of the position of the maximum $\mathrm{q \left( I_{MAX} \right)}$, of the half width $\mathrm{\Delta L_{FSDP}}$ of the FSDP and the intensity extrapolated at $\mathrm{q \ = \ 0 \ nm^{-1}}$ $\mathrm{I \left( q = 0 \right) }$ for the Belt 9.2\% and Belt 15.5\% samples annealed at 770 \degree C. The different parameters are expressed as a function of the temperature during the heating part, and as a function of time during the isothermal annealing. The black points correspond to the temperatures at which a maximum is visible in the area of the D2 band in the Raman spectrum for the corresponding sample.}
\label{fig:DRX_Belt_15_9}
\end{figure}

In-situ evolution of the FSDP during an annealing at 770 \degree C during 6.5 hours for the Belt 16.5\% is shown in the figure \ref{fig:FSDP_example}.
The FSDP of the densified glass at room temperature corresponds to the curve with a maximum at the highest q. During the heating ramp, the FSDP shifts slightly to lower q, down to the red spectrum, which corresponds to the first spectrum acquired at the annealing temperature. During the heat treatment, the position evolves continuously toward the position of the non-densified glass, marked by an arrow in the figure \ref{fig:FSDP_example}. This means that the correlation at the origin of the FSDP take place over greater distances as the relaxation goes on. Note in figure \ref{fig:FSDP_example} that as the FSDP shifts to low q, its tail ranges out of the detection window. Because of the experimental setup, it was impossible to move the WAXS detector further, and to get the whole FSDP. To avoid any underestimation of the FSDP width, the peaks were fitted with Pearson IV distributions, giving $\mathrm{\Delta L_{FSDP}}$ (see SI for complete process). $\mathrm{I \left( q = 0 \right) }$ is extrapolated by a Guinier analysis.\newline
Annealings were performed at 770\degree C for the Belt 15.5\% and Belt 9.2\% and at 900 \degree C for the Belt 16.5\%. The results, i.e. the FSDP position $\mathrm{q \left( I_{MAX} \right)}$ and width $\mathrm{\Delta L_{FSDP}}$, and the extrapolation of the intensity at zero scattering angle $\mathrm{I \left( q = 0 \right) }$  are shown in figure \ref{fig:DRX_Belt_16} and \ref{fig:DRX_Belt_15_9}.
The missing data in the figure \ref{fig:DRX_Belt_16} are due to a loss of the beam during the experiment. The dashed lines, corresponding to the results of other runs (the run of the figure \ref{fig:FSDP_example} for the WAXS data) complete the data sets.
The position of the maximum of the FSDP behaves similarly for all samples. The initial plateaus indicate that very modest changes appear at low temperature, followed by monotoneous decrease after different thresholds: above 800, 600 and 200\degree C for the Belt 16.5\%, 15.5\% and 9.2\%. The evolution of $\mathrm{q \left( I_{MAX} \right)}$ is very smooth throughout the transformation, and no discontinuity or sudden change of rate is visible for any of the initial densification ratio. At the end of the relaxation, the positions of the FSDP,  are close to these of  pristine silica glass, $\mathrm{15.1 \geq q \left( I_{MAX} \right) \geq 14.9 \ nm}$ (SI), indicating a completion of roughly 80\% of the transformation.

On the other hand, the evolution of the FSDP width differs strongly between the different samples during the experiments. For the sample Belt 16.5\%, one observes a slight decrease for a wide range of temperature, followed by a sudden increase initiating around 800 \degree C. This increase continues during the isothermal annealing, to reach a maximum when $\mathrm{t = 600 \ s}$. $\mathrm{\Delta L_{FSDP}}$ then decreases continuously from that maximum. For the samples Belt 15.5\% and Belt 9.2\%, a general decrease is observed as the temperature increases from room temperature, a decrease that starts immediately as temperature rises. At high temperature, around 720\degree C (Belt 9.2\%) and 770\degree C (Belt 15.5\%), there is a local maximum in $\mathrm{\Delta L_{FSDP}}$ followed by a continuous decrease during the isothermal annealing.
For pristine silica at 770\degree C, $\mathrm{\Delta L_{FSDP} = 5.4 \pm 0.05 nm^{-1}}$. At 900\degree C it can be estimated at $\mathrm{\Delta L_{FSDP} = 5.3 \pm 0.05 nm^{-1}}$ (SI). The final value at the end of the experiments shown in figure \ref{fig:DRX_Belt_16} and \ref{fig:DRX_Belt_15_9} are 5.97, 5.64 and \SI{5.88}{\per\nano\metre} ($\mathrm{\pm 0.03}$) for the samples Belt 16.5\%, 15.5\% and 9.2\% respectively.\newline
The SAXS data show a unique maximum in $\mathrm{I \left( q = 0 \right) }$ for the three different samples, and this maximum is much more pronounced than for $\mathrm{\Delta L_{FSDP}}$.
The maxima in $\mathrm{I \left( q = 0 \right) }$ and $\mathrm{\Delta L_{FSDP}}$ appear during the heating ramp for Belt 9.2\% and 15.5\% samples, and during the isothermal annealing for highly densified Belt 16.5\% glass. These maxima coincide only for Belt 16.5\% glass. Otherwise the maximum in $\mathrm{I \left( q = 0 \right) }$ appears at lower temperatures with respect to the maximum in $\mathrm{\Delta L_{FSDP}}$. Furthermore, for the sample Belt 16.5\% only, $\mathrm{I \left( q = 0 \right) }$ decreases for a significant part of the heating, indicating a reduction of the density fluctuations at this stage. Additionally, the curves that represent $\mathrm{I \left( q = 0 \right) }$ and $\mathrm{\Delta L_{FSDP}}$ are very much alike. The structural orders underlying the FSDP and the SAXS intensity have similar relaxation dynamics for this Belt 16.5\% sample.\newline
As adressed in our previous paper, a transitory disordered state is visible during the relaxation through a maximum in the area of the so-called D2 band in the Raman spectrum. This D2 band is related to the rings made of 3 $\mathrm{SiO_{4}}$ tetrahedra only\cite{Cornet2017}, so the disorder probed in Raman is on the length scale of the tetrahedron rings within the silica network, which is a shorter length scale that the one related to $\mathrm{\Delta L_{FSDP}}$. Therefore, to compare the dynamics of each length scale and the structural orders associated, we represent here our Raman results, i.e. the maximum of the D2 band area, comparatively to the signal evolution in SAXS and WAXS. For the samples Belt 9.2\% and Belt 15.5\%, where the maximum in $\mathrm{\Delta L_{FSDP}}$ and $\mathrm{I(q=0)}$ appears during the heating ramp, the temperatures corresponding to the D2 maximum are reported in the figure \ref{fig:DRX_Belt_15_9} by black points. For these samples, the maximum in the D2 band area appears at the same temperatures than the maximum in $\mathrm{\Delta L_{FSDP}}$.
\begin{figure}[h!]
\center
\includegraphics[width=9cm]{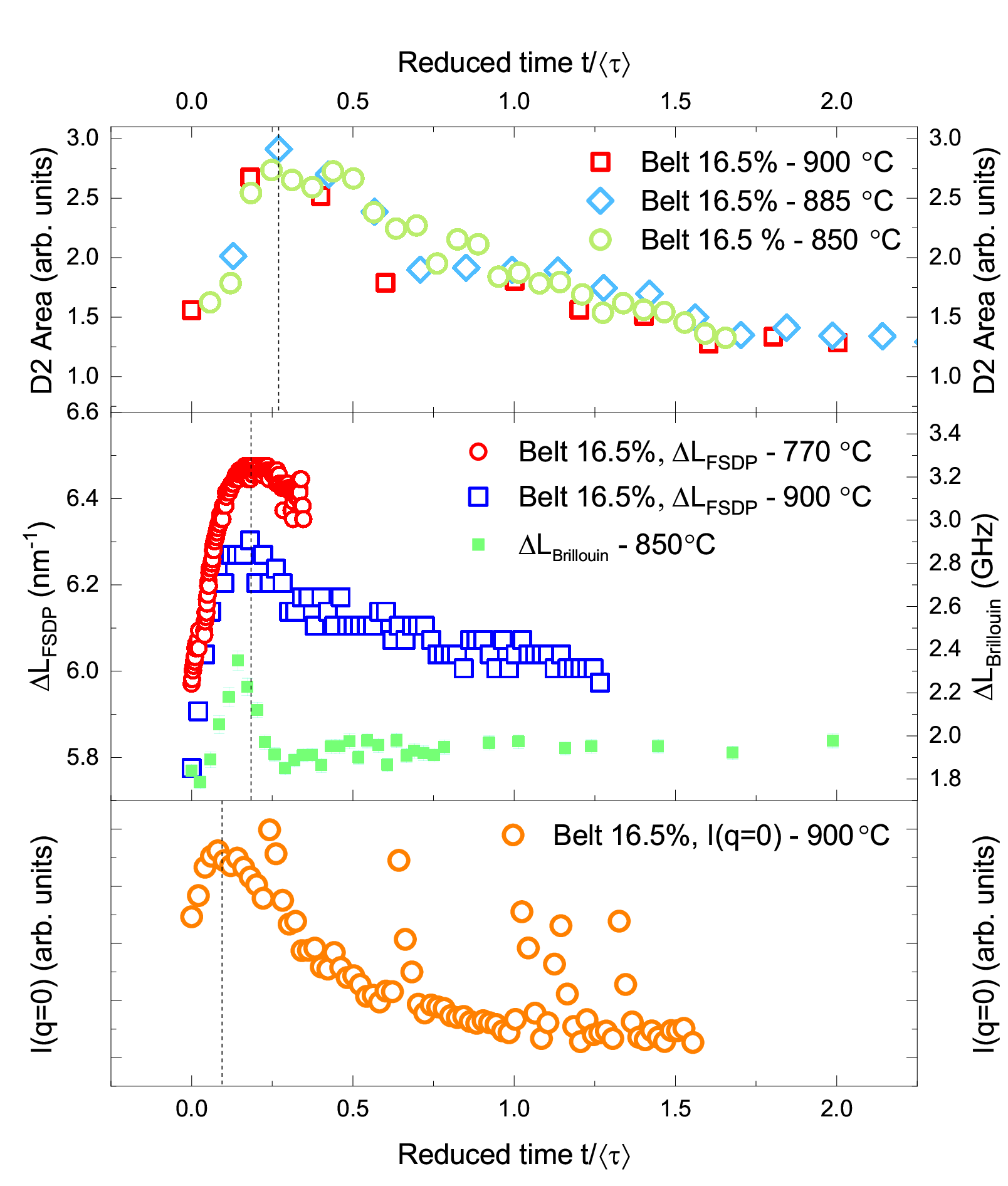}
\caption{Normalized D2 band area (from Ref.\onlinecite{Cornet2017}), $\mathrm{\Delta L_{FSDP}}$, $\mathrm{I \left( q = 0 \right) }$ as a function of the reduced time $t / \langle \tau \rangle$. The curve of $\mathrm{\Delta L_{FSDP}}$ at 770\degree C is shifted vertically for clarity. $\mathrm{\Delta L_{Brillouin}}$, hwhm of the longitudinal Brillouin peak obtained by Guerette \textit{et. al.} during the relaxation at 850\degree C of a sample densified at 4 GPa and 1100\degree C is added in the middle panel\cite{Guerette2018}.}
\label{fig:D1020_D2_DRX}
\end{figure}
For the Belt 16.5\% sample, a direct comparison of the annealings performed with Raman and WAXS spectroscopies is shown in the figure \ref{fig:D1020_D2_DRX}.

In this figure, we plotted the data as a function of a reduced time $t / \langle \tau \rangle$. The characteristic times of the relaxation $\langle \tau \rangle$ are those obtained from Ref. \onlinecite{Cornet2017}, where annealings were performed at different temperatures from 500\degree C to 900\degree C for each Belt sample. Different annealing temperatures need to be chosen in such a way that the whole transformation can be observed within reasonable experimental times, i.e. within days. It was found that the relaxation of the main structural features in the Raman spectra can be modelled effectively by stretched exponential, and that the characteristic times from all the samples independently of their initial densification have a similar dependence with annealing temperature, following the expression: $\langle \tau \rangle^{-1} = A \times exp(- E_{a} / RT )$ ($\langle \tau \rangle$ in seconds, T in Kelvin, $E_{a} = \SI{255.8}{\kilo\joule\per\mole}$, R= \SI{8.314}{\joule\per\kelvin\per\mol} and A = \SI{81.82e6}{\per\second}). Dividing the time of an isotherm by its characteristic time allows comparing experiment realized at different temperatures. However, it is not sure that the processes observed in x-ray scattering and Raman spectroscopy have the same characteristic times $\langle \tau \rangle$. In the figure \ref{fig:D1020_D2_DRX} is shown the Belt 16.5\% sample $\mathrm{\Delta L_{FSDP}}$ data for two isotherms at 770\degree C and 900\degree C rescaled using the characteristic times determined by Raman spectroscopy. The consistent scaling of the two data sets justifies the use on the X-ray data of the characteristic relaxation times obtained from Raman spectroscopy.

The maximum in the area of the D2 band area is witnessed at a similar relaxation time that the maximum in $\mathrm{\Delta L _{FSDP}}$. Therefore, the maxima observed in the area of the Raman D2 band and in the width of the FSDP coincide during the transformation for the three initial densification ratio.

Finally, the data obtained by Guerette \textit{et. al.} with Brillouin Light Scattering show evidence of a similar result at a larger scale, corresponding to the continuum approach \cite{Guerette2018}. We plotted the data of Guerette, i.e. the evolution of the half width at half maximum of the longitudinal Brillouin peak $\mathrm{\Delta L _{Brillouin}}$ during the relaxation at 850\degree C of a silica sample densified at 4 GPa and 1100\degree C in the lower panel of the figure \ref{fig:D1020_D2_DRX}, assuming that the expression for $\langle \tau \rangle$ holds for these compression parameters. We observe that the maxima in $\mathrm{\Delta L _{Brillouin}}$ and in $\mathrm{\Delta L _{FSDP}}$ are simultaneous.

\section{\label{sec:level4}Discussion}

It is the first time in the literature that a structural study of the temperature relaxation process of densified glass has been observed with x-ray scattering. We will now discuss the results for each length scale, i.e. for the FSDP then for the SAXS data. In a last part, we will discuss the dynamics of each structural feature comparatively, to describe the relaxation processes in a unified picture, from a few interatomic distances up to the continuum scale.\\

Qualitative observation of the FSDP behavior during the relaxation has been done above from the figure \ref{fig:FSDP_example}. The primary observation is the shift to lower scattering vector q. From the general interpretation of the FSDP, it means that the caracteristic distance of the IRO (Intermediate Range Order) increases continously. Although the FSDP cannot be linked directly to the density \cite{Bauchy2012}, a general trend exists \cite{Sokolov1992, Zanatta2014, Salmon2018}, and the shift to low q is consistent with the monotonous decrease in the density that we observe (SI). Quantitative evolution of the position of the FSDP is available in the figures \ref{fig:DRX_Belt_16} and \ref{fig:DRX_Belt_15_9}. As seen in our previous Raman study, the lower the temperature during densification, the less stable the densification and lower temperature the relaxation takes place. This can be seen by the plateaus in the evolution $\mathrm{q \left( I_{MAX} \right) }$, that end  above 800, 600 and 200\degree C for the Belt 16.5, 15.5 and 9.2\%. Interestingly, the transition temperature during the relaxation is known to decrease with increasing densification pressure \cite{Guerette2018}.

Nevertheless, the decrease in density visible through the evolution of the FSDP does not prevent any specific increase in the fluctuations, that should be visible in its width $\mathrm{\Delta L_{FSDP}}$. The general trend in the behavior of $\mathrm{\Delta L_{FSDP}}$, global increase or decrease, depends on the initial value of $\mathrm{\Delta L_{FSDP}}$. This initial value depends on the densification process, as higher temperature during the compression produces a more homogeneous glass \cite{Guerette2015, Cornet2017, Guerette2018}, leading to lower $\mathrm{\Delta L_{FSDP}}$ values for densified samples, eventually lower than $\mathrm{\Delta L_{FSDP}}$ for pristine silica glass ($\mathrm{5.6 \ nm^{-1}}$) for the Belt 16.5\%. However, there is a (local) maximum in the evolution of $\mathrm{\Delta L_{FSDP}}$ during the temperature relaxation process for all the tested samples. Recalling that the distance over which the ring structure in the intermediate range order (IRO) of the glass remains coherent is given by $\mathrm{L = 2 \pi / \Delta L_{FSDP}}$, the maximum in $\mathrm{\Delta L_{FSDP}}$ indicates the presence of higher disorder in the IRO. This is the confirmation that the relaxation in IRO goes through a transitory state.

During the relaxation, the values of $\mathrm{\Delta L_{FSDP}}$ at the local maxima are 6.27, 6.03 and 6.21 $\mathrm{nm^{-1}}$ for the samples Belt 16.5\%, 15.5\% and 9.2\% respectively. The magnitude of the maximum in $\mathrm{\Delta L_{FSDP}}$, i.e. the difference between these values and these of pristine silica (SI) are $\mathrm{0.99 \pm 0.12}$, $\mathrm{0.67 \pm 0.06}$ and $\mathrm{0.85 \pm 0.08 \ nm^{-1}}$. These values are somewhat close, compared to the total evolution of $\mathrm{\Delta L_{FSDP}}$, but not enough to assess that the increase in the inhomogeneities during the relaxation is the same regardless the initial structure. In that, the relaxation differ from the observation made on the Raman D2 band only. Therefore the relaxation processes are different at different (but close) length scales. No scaling law can thus be deduced from our results. Finally, $\mathrm{\Delta L_{FSDP}}$ converges very slowly to the pristine values, at a rate that is not compatible with reasonnable observation times. This contrasts with the observation of a near complete relaxation process indicated by $\mathrm{q \left( I_{MAX} \right)}$.\\

The intensity of the SAXS diffusion at the zero scattering angle $\mathrm{I(q=0)}$ evolves through a maximum for each annealing, showing a state of maximum density fluctuations. The maxima are much more pronounced than for $\mathrm{\Delta L_{FSDP}}$. For pristine silica and in this range of temperature, the evolution of $\mathrm{I(q=0)}$ is due to changes in the dynamic fluctuations induced by the acoustic phonons (see eq \ref{eq:SAXS_density_bis}). For all samples, we know that the density decreases monotonously during the relaxation process (SI). Formally, $V_{l,\infty}$ should be derived using an high frequency probe, e.g. Inelastic X-Ray Scattering (IXS). However, in densified silica, a linear dispersion relation has been shown to hold from the gigahertz to the terahertz domains \cite{Zanatta2010}, thus it is possible to derive $V_{l,\infty}$ from Brillouin Light Scattering (BLS) measurements. Such BLS experiments were recently performed by Guerette \textit{et. al.} for a silica sample densified at 4 GPa and 1100\degree C. The longitudinal frequency shift during an isothermal annealing at 850\degree C shows a monotonous decrease, while the transition is visible in the peak width \cite{Guerette2018}. In their experiment, the shift correlates directly to $V_{l,\infty}$. Thus, the density $\rho$ and the longitudinal sound wave velocity $V_{l,\infty}$ both decrease monotonously with temperature, and the dynamic part in Eq. \ref{eq:SAXS_density_bis} increases \emph{continuously}. The pronounced maximum observed here is then related to an evolution of the static part of the density fluctuations. Thus, this maximum that we observe is another evidence that a polyamorphic transition occurs, through a transitory state characterized by a maximum of inhomogeneities, here on the density fluctuations over a length scale of about 1-5 nm.\\

\begin{figure}[h!]
\center
\includegraphics[width=9cm]{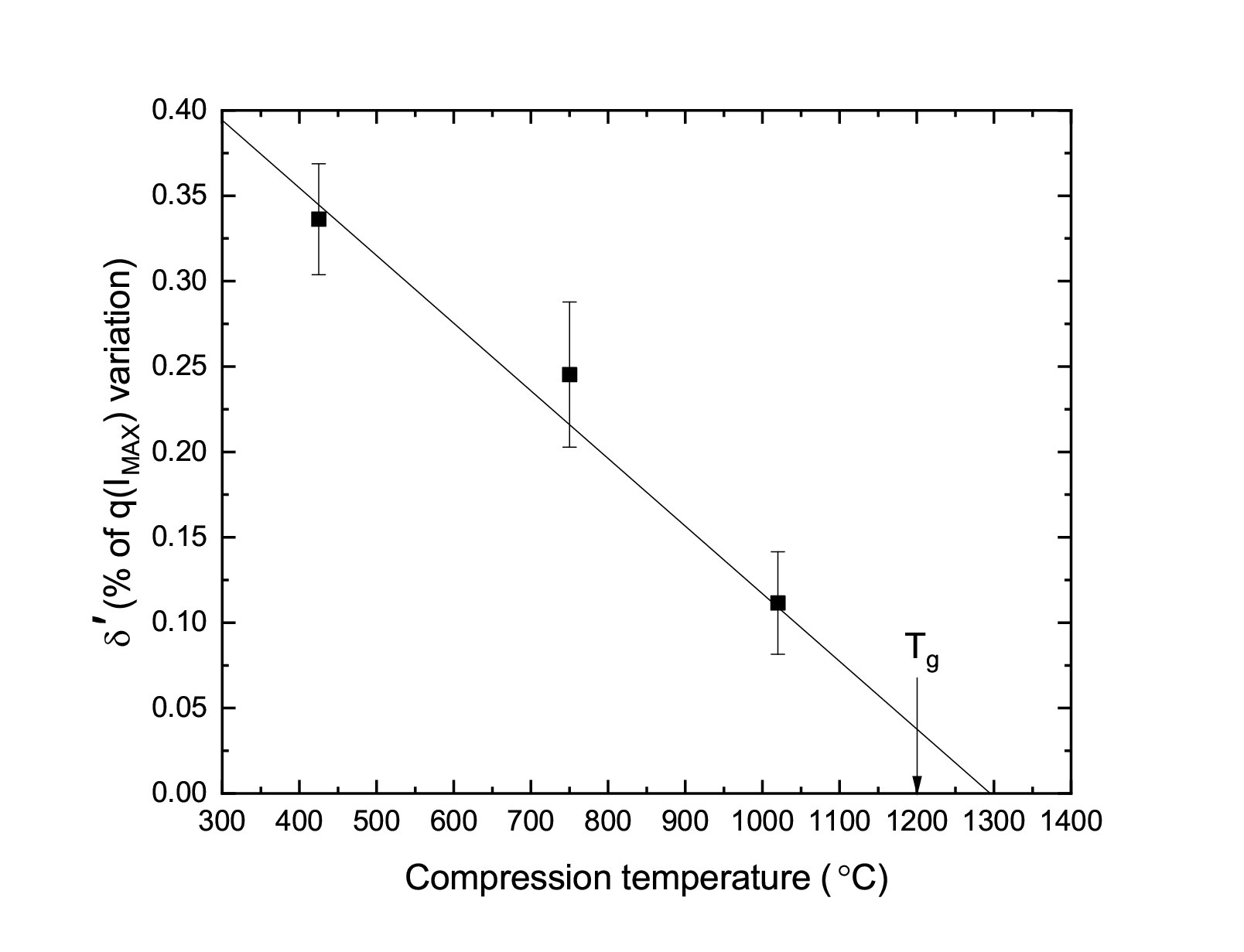}
\caption{Discrepancy $\mathrm{\delta}$ between $\mathrm{\Delta L_{FSDP}}$ and $\mathrm{I(q=0)}$ maxima, as a function of the densification temperature. The value of delta at 1020\degree C is an upper bound, as represented by the arrow, and the complete simultaneity at $\mathrm{T_{g}}$ is hypothetical, and not experimentally derived. The dashed line is a guide for the eyes.}
\label{fig:delta}
\end{figure}

The maxima observed in $\mathrm{\Delta L_{FSDP}}$ and $\mathrm{I(q=0)}$ are not exactly concomitant. It can be seen in figures \ref{fig:DRX_Belt_16} and \ref{fig:DRX_Belt_15_9} that the maximum in $\mathrm{I(q=0)}$ appears before the maximum in the FSDP width during the relaxation process, with a gap $\mathrm{\delta}$ that depends on the densification conditions. The two maxima are observed almost simultaneously for Belt 16.5\% and are very distinct for Belt 9.2\%. Depending on the experiment the transformation takes place at different time and temperature and it is ambiguous to quantify the gaps $\mathrm{\delta}$ to compare them. However the gap $\mathrm{\delta}$ can be scaled by the corresponding relative variation of $\mathrm{q \left( I_{MAX} \right)}$, namely $\mathrm{\delta '}$. The values $\mathrm{q \left( I_{MAX} \right)}$ at room temperature before the annealing and for pristine silica at the annealing temperature define the initial and final values for the relaxation of $\mathrm{q \left( I_{MAX} \right)}$. If the maximum of the $\mathrm{I(q=0)}$ coincide approximately to the inflection point of the variation of $\mathrm{q \left( I_{MAX} \right)}$ around 40\% of completion for all samples, the maximum of the $\mathrm{\Delta L_{FSDP}}$ shifts toward the fully completion of the $\mathrm{I(q=0)}$ evolution when the sample density decreases (lower compression temperature). For the Belt 16.5\% sample, the gap $\mathrm{\delta}$ is equivalent to a variation of $\mathrm{I(q=0)}$ by only $\mathrm{ \delta ' = 12 \pm 5 \%}$ when it is up to $\mathrm{ \delta ' = 12 \pm 5 \%}$ for Belt 9.2\% sample. This evolution reflects well the homogeneity of the recovered densified sample. It was found that lower is the compression temperature with an identical densification pressure (5 GPa), more heterogeneous is the sample. The gap $\mathrm{\delta}$ (or $\mathrm{\delta '}$) can then be interpreted as a decoupling between relaxation at the mid-range order (signal of $\mathrm{I(q=0)}$) or at the intermediate range order (signal of $\mathrm{\Delta L_{FSDP}}$). The dynamics of the different length scale differ in relation to the heterogeneities. To test further this idea in figure \ref{fig:delta} is plotted the gap $\mathrm{\delta '}$ in function of the compression temperature. The values correlate very well. If a linear extrapolation is made it is then found that $\mathrm{\delta '}$ will be zero for a temperature of $\mathrm{1290\pm100 \degree C}$ so very close to the nominal $\mathrm{T_{g}}$ of $\mathrm{SiO_{2}}$ which is around 1200\degree C. The higher is the temperature during densification, narrower is the distribution of activation energy and the densified glass reacts at the same time at all scales.

Unfortunately, we are still lacking an comprehensive description of the intermediate range order (IRO) of (silica) glass. Moreover, the influence of the temperature on the IRO in compression remains elusive. Indeed, due to the combination of challenging experiments (\textit{in situ} HP-HT) and weak or hard-to-extract spectroscopic features (boson peak, $\mathrm{I(q=0)}$), no complete systematic study of the IRO with respect to temperature in compression is available to our knowledge (only the FSDP for $\mathrm{SiO_{2}}$ and $\mathrm{GeO_{2}}$ glasses \cite{Inamura2004, Shen2007}).\\

Nevertheless, we can draw a parallel with cold densified silica glass to discuss the SAXS/WAXS decoupling. At ambient temperature, permament densification is achieved in silica glass when the pressure goes beyond 9 GPa, the elastic limit. The recovered sample possesses a complex structure, where the low (LDA, Low Density Amorphous) and high density (HDA, High Density Amorphous) polyamorphic states are intricated. Pure HDA glass is obtained only if the compression pressure is high enough (about 25 GPa) \cite{Sonneville2012}.
Similarly, a silica sample compressed at a certain pressure will have a particular ratio of HDA and LDA states, depending on temperature. It is important to note that this HDA state is not the same as the HDA state obtained at room temperature, but only temperatures close to the glass transition temperature $T_{g}$ allow complete densification, and the recovery of a complete HDA glass. Given the glass transition temperature of suprasil 300: $T_{g}$ = 1200\degree C, only the sample Belt 16.5\%, compressed at 1020\degree C, can be considered as nearly pure HDA. Consequently, only the relaxation of this sample is a pure HDA-LDA transition. As the behaviour of $\mathrm{\delta '}$ suggest, we observe precisely that the maxima observed in $\mathrm{I(q=0)}$ and $\mathrm{\Delta L_{FSDP}}$ coincide more and more as the compression temperature of the sample tend to $T_{g}$, with near simultaneity for the sample Belt 16.5\%. Our proposed interpretation is that, for a pure HDA-LDA transition at high temperature, the maxima in structural fluctuations observed at the length scales corresponding to the Raman D2 band, the FSDP and the density fluctuations are simultaneous.\newline
The picture of an HDA-LDA transition through an activated transition state characterized by a maximum of heterogeneities at all scales is coherent with the current understanding of polyamorphism in water, where a pure HDA-LDA transition is achieved. The HDA-LDA transition in amorphous water ice was shown to imply a State of Strongest Heterogeneity (SSH) in the nanometer scale, corresponding to the maximum in $\mathrm{I(q=0)}$ described in this study. Furthermore, this SSH is also simultaneous with $\mathrm{\Delta L_{FSDP}}$ during the transformation\cite{Koza2006, Koza2007}.
Kim et. al. documented the liquid polyamorphic transition between the Low Density and High Density Liquids (LDL-HDL), in the deeply supercooled domain (228K in vacuum), but still at the thermodynamic equilibrium. They found a perfect simultaneity between the heterogeneity probed at the SAXS and WAXS length scales\cite{Kim2017}. These heterogeneities or SSH states observed in water are similar to the transitory state here observed in the case of $\mathrm{\delta ' = 0}$ we could have achieved if the densification have been done closer to $\mathrm{T_{g}}$.

\section{\label{sec:level5}Conclusion}

A transitory state was evidenced in the structure of the silica glass during relaxation of densified samples at high temperature. This transitory state is defined by a maximum in the structural heterogeneities. We propose that this activated transitory state between the two more ordered high and low density amorphous states confirms the existence of two mega basin in the energy landscape and therefore an amorphous-amorphous transition.  With a combination of experimental probes, namely small and wide x-ray scattering, and in combination with previous Raman results, we confirm that this transitory state appears at the different length scales from a few interatomic distances to several nanometers. The dynamics of the relaxation processes at the length scales covered by the previous results in Raman and the FSDP width are the same whatever the compression temperature of the samples. When a pure HDA-LDA transition is monitored, i.e. if the glass had been densified close to the glass transition temperature, the transitory state is observed simultaneously at all length scales, from few interatomic bonds to several nanometers, eventually up to the continuum approach. When the densification is not complete, a more complex relaxation at the density fluctuation scale needs to be further elucidated. This constitutes the first multiscale evidence of polyamorphism in oxide glasses, and opens the gates to a more detailed statistical analysis of the phenomenon.\\

\section{Supplementary Material}

Additional content is available in the supplementary material. This covers the processing of the scattering data, the first sharp diffraction peak evolution with temperature in pristine silica as well the necessary data to establish the connection between our previous results from Raman spectroscopy to the Raman data presented in the figure \ref{fig:DRX_Belt_16}, \ref{fig:DRX_Belt_15_9} and \ref{fig:D1020_D2_DRX}.

\begin{acknowledgments}
The authors wish to thank the whole team in the BM26 beamline in the ESRF, especially Daniel Hermida Merino. The compression of the samples were realized in the PLECE plateform in Lyon, with Sylvie Le Floch, whose help was greatly appreciated. The authors declare no conflict of interest.
\end{acknowledgments}

\bibliography{biblio}

\providecommand{\noopsort}[1]{}\providecommand{\singleletter}[1]{#1}%
\begin{thebibliography}{48}%
\makeatletter
\providecommand \@ifxundefined [1]{%
 \@ifx{#1\undefined}
}%
\providecommand \@ifnum [1]{%
 \ifnum #1\expandafter \@firstoftwo
 \else \expandafter \@secondoftwo
 \fi
}%
\providecommand \@ifx [1]{%
 \ifx #1\expandafter \@firstoftwo
 \else \expandafter \@secondoftwo
 \fi
}%
\providecommand \natexlab [1]{#1}%
\providecommand \enquote  [1]{``#1''}%
\providecommand \bibnamefont  [1]{#1}%
\providecommand \bibfnamefont [1]{#1}%
\providecommand \citenamefont [1]{#1}%
\providecommand \href@noop [0]{\@secondoftwo}%
\providecommand \href [0]{\begingroup \@sanitize@url \@href}%
\providecommand \@href[1]{\@@startlink{#1}\@@href}%
\providecommand \@@href[1]{\endgroup#1\@@endlink}%
\providecommand \@sanitize@url [0]{\catcode `\\12\catcode `\$12\catcode
  `\&12\catcode `\#12\catcode `\^12\catcode `\_12\catcode `\%12\relax}%
\providecommand \@@startlink[1]{}%
\providecommand \@@endlink[0]{}%
\providecommand \url  [0]{\begingroup\@sanitize@url \@url }%
\providecommand \@url [1]{\endgroup\@href {#1}{\urlprefix }}%
\providecommand \urlprefix  [0]{URL }%
\providecommand \Eprint [0]{\href }%
\providecommand \doibase [0]{http://dx.doi.org/}%
\providecommand \selectlanguage [0]{\@gobble}%
\providecommand \bibinfo  [0]{\@secondoftwo}%
\providecommand \bibfield  [0]{\@secondoftwo}%
\providecommand \translation [1]{[#1]}%
\providecommand \BibitemOpen [0]{}%
\providecommand \bibitemStop [0]{}%
\providecommand \bibitemNoStop [0]{.\EOS\space}%
\providecommand \EOS [0]{\spacefactor3000\relax}%
\providecommand \BibitemShut  [1]{\csname bibitem#1\endcsname}%
\let\auto@bib@innerbib\@empty
\bibitem [{\citenamefont {Rapoport}(1967)}]{Rapoport1967}%
  \BibitemOpen
  \bibfield  {author} {\bibinfo {author} {\bibfnamefont {E.}~\bibnamefont
  {Rapoport}},\ }\bibfield  {title} {\enquote {\bibinfo {title} {Model for
  melting‐curve maxima at high pressure},}\ }\href {\doibase
  10.1063/1.1841150} {\bibfield  {journal} {\bibinfo  {journal} {The Journal of
  Chemical Physics}\ }\textbf {\bibinfo {volume} {46}},\ \bibinfo {pages}
  {2891--2895} (\bibinfo {year} {1967})}\BibitemShut {NoStop}%
\bibitem [{\citenamefont {Mishima}, \citenamefont {Calvert},\ and\
  \citenamefont {Whalley}(1985)}]{Mishima1985}%
  \BibitemOpen
  \bibfield  {author} {\bibinfo {author} {\bibfnamefont {O.}~\bibnamefont
  {Mishima}}, \bibinfo {author} {\bibfnamefont {L.}~\bibnamefont {Calvert}}, \
  and\ \bibinfo {author} {\bibfnamefont {E.}~\bibnamefont {Whalley}},\
  }\bibfield  {title} {\enquote {\bibinfo {title} {An apparently first-order
  transition between two amorphous phases of ice induced by pressure},}\ }\href
  {\doibase 10.1038/314076a0} {\bibfield  {journal} {\bibinfo  {journal}
  {Nature}\ }\textbf {\bibinfo {volume} {314}},\ \bibinfo {pages} {76--78}
  (\bibinfo {year} {1985})}\BibitemShut {NoStop}%
\bibitem [{\citenamefont {Loerting}\ \emph {et~al.}(2011)\citenamefont
  {Loerting}, \citenamefont {Winkel}, \citenamefont {Seidl}, \citenamefont
  {Bauer}, \citenamefont {Mitterdorfer}, \citenamefont {Handle}, \citenamefont
  {Salzmann}, \citenamefont {Mayer}, \citenamefont {Finney},\ and\
  \citenamefont {Bowron}}]{Loerting2011}%
  \BibitemOpen
  \bibfield  {author} {\bibinfo {author} {\bibfnamefont {T.}~\bibnamefont
  {Loerting}}, \bibinfo {author} {\bibfnamefont {K.}~\bibnamefont {Winkel}},
  \bibinfo {author} {\bibfnamefont {M.}~\bibnamefont {Seidl}}, \bibinfo
  {author} {\bibfnamefont {M.}~\bibnamefont {Bauer}}, \bibinfo {author}
  {\bibfnamefont {C.}~\bibnamefont {Mitterdorfer}}, \bibinfo {author}
  {\bibfnamefont {P.~H.}\ \bibnamefont {Handle}}, \bibinfo {author}
  {\bibfnamefont {C.~G.}\ \bibnamefont {Salzmann}}, \bibinfo {author}
  {\bibfnamefont {E.}~\bibnamefont {Mayer}}, \bibinfo {author} {\bibfnamefont
  {J.~L.}\ \bibnamefont {Finney}}, \ and\ \bibinfo {author} {\bibfnamefont
  {D.~T.}\ \bibnamefont {Bowron}},\ }\bibfield  {title} {\enquote {\bibinfo
  {title} {How many amorphous ices are there?}}\ }\href {\doibase
  10.1039/c0cp02600j} {\bibfield  {journal} {\bibinfo  {journal} {Physical
  Chemistry Chemical Physics}\ }\textbf {\bibinfo {volume} {13}},\ \bibinfo
  {pages} {8783--8794} (\bibinfo {year} {2011})}\BibitemShut {NoStop}%
\bibitem [{\citenamefont {Brazhkin}\ and\ \citenamefont
  {Lyapin}(2003)}]{Brazhkin2003}%
  \BibitemOpen
  \bibfield  {author} {\bibinfo {author} {\bibfnamefont {V.~V.}\ \bibnamefont
  {Brazhkin}}\ and\ \bibinfo {author} {\bibfnamefont {A.~G.}\ \bibnamefont
  {Lyapin}},\ }\bibfield  {title} {\enquote {\bibinfo {title} {High-pressure
  phase transformations in liquids and amorphous solids},}\ }\href
  {http://stacks.iop.org/0953-8984/15/i=36/a=301} {\bibfield  {journal}
  {\bibinfo  {journal} {Journal of Physics: Condensed Matter}\ }\textbf
  {\bibinfo {volume} {15}},\ \bibinfo {pages} {6059} (\bibinfo {year}
  {2003})}\BibitemShut {NoStop}%
\bibitem [{\citenamefont {McMillan}(2004)}]{McMillan2004}%
  \BibitemOpen
  \bibfield  {author} {\bibinfo {author} {\bibfnamefont {P.}~\bibnamefont
  {McMillan}},\ }\bibfield  {title} {\enquote {\bibinfo {title} {Polyamorphic
  transformations in liquids and glasses},}\ }\href {\doibase 10.1039/b401308p}
  {\bibfield  {journal} {\bibinfo  {journal} {Journal of Materials Chemistry}\
  }\textbf {\bibinfo {volume} {14}},\ \bibinfo {pages} {1506--1512} (\bibinfo
  {year} {2004})}\BibitemShut {NoStop}%
\bibitem [{\citenamefont {Wilding}, \citenamefont {Wilson},\ and\ \citenamefont
  {McMillan}(2006)}]{Wilding2006}%
  \BibitemOpen
  \bibfield  {author} {\bibinfo {author} {\bibfnamefont {M.~C.}\ \bibnamefont
  {Wilding}}, \bibinfo {author} {\bibfnamefont {M.}~\bibnamefont {Wilson}}, \
  and\ \bibinfo {author} {\bibfnamefont {P.~F.}\ \bibnamefont {McMillan}},\
  }\bibfield  {title} {\enquote {\bibinfo {title} {Structural studies and
  polymorphism in amorphous solids and liquids at high pressure},}\ }\href
  {\doibase 10.1039/b517775h} {\bibfield  {journal} {\bibinfo  {journal}
  {Chemical Society Reviews}\ }\textbf {\bibinfo {volume} {35}},\ \bibinfo
  {pages} {964--986} (\bibinfo {year} {2006})}\BibitemShut {NoStop}%
\bibitem [{\citenamefont {Machon}\ \emph {et~al.}(2014)\citenamefont {Machon},
  \citenamefont {Meersman}, \citenamefont {Wilding}, \citenamefont {Wilson},\
  and\ \citenamefont {McMillan}}]{Machon2014}%
  \BibitemOpen
  \bibfield  {author} {\bibinfo {author} {\bibfnamefont {D.}~\bibnamefont
  {Machon}}, \bibinfo {author} {\bibfnamefont {F.}~\bibnamefont {Meersman}},
  \bibinfo {author} {\bibfnamefont {M.}~\bibnamefont {Wilding}}, \bibinfo
  {author} {\bibfnamefont {M.}~\bibnamefont {Wilson}}, \ and\ \bibinfo {author}
  {\bibfnamefont {P.}~\bibnamefont {McMillan}},\ }\bibfield  {title} {\enquote
  {\bibinfo {title} {Pressure-induced amorphization and polyamorphism:
  Inorganic and biochemical systems},}\ }\href {\doibase
  https://doi.org/10.1016/j.pmatsci.2013.12.002} {\bibfield  {journal}
  {\bibinfo  {journal} {Progress in Materials Science}\ }\textbf {\bibinfo
  {volume} {61}},\ \bibinfo {pages} {216 -- 282} (\bibinfo {year}
  {2014})}\BibitemShut {NoStop}%
\bibitem [{\citenamefont {Morishita}(2004)}]{Tetsuya2004}%
  \BibitemOpen
  \bibfield  {author} {\bibinfo {author} {\bibfnamefont {T.}~\bibnamefont
  {Morishita}},\ }\bibfield  {title} {\enquote {\bibinfo {title} {High density
  amorphous form and polyamorphic transformations of silicon},}\ }\href
  {\doibase 10.1103/PhysRevLett.93.055503} {\bibfield  {journal} {\bibinfo
  {journal} {Phys. Rev. Lett.}\ }\textbf {\bibinfo {volume} {93}},\ \bibinfo
  {pages} {055503} (\bibinfo {year} {2004})}\BibitemShut {NoStop}%
\bibitem [{\citenamefont {Liang}, \citenamefont {Miranda},\ and\ \citenamefont
  {Scandolo}(2007)}]{Scandolo2007}%
  \BibitemOpen
  \bibfield  {author} {\bibinfo {author} {\bibfnamefont {Y.}~\bibnamefont
  {Liang}}, \bibinfo {author} {\bibfnamefont {C.~R.}\ \bibnamefont {Miranda}},
  \ and\ \bibinfo {author} {\bibfnamefont {S.}~\bibnamefont {Scandolo}},\
  }\bibfield  {title} {\enquote {\bibinfo {title} {Mechanical strength and
  coordination defects in compressed silica glass: Molecular dynamics
  simulations},}\ }\href {\doibase 10.1103/PhysRevB.75.024205} {\bibfield
  {journal} {\bibinfo  {journal} {Phys. Rev. B}\ }\textbf {\bibinfo {volume}
  {75}},\ \bibinfo {pages} {024205} (\bibinfo {year} {2007})}\BibitemShut
  {NoStop}%
\bibitem [{\citenamefont {Benmore}\ \emph {et~al.}(2010)\citenamefont
  {Benmore}, \citenamefont {Soignard}, \citenamefont {Amin}, \citenamefont
  {Guthrie}, \citenamefont {Shastri}, \citenamefont {Lee},\ and\ \citenamefont
  {Yarger}}]{Yarger2010}%
  \BibitemOpen
  \bibfield  {author} {\bibinfo {author} {\bibfnamefont {C.~J.}\ \bibnamefont
  {Benmore}}, \bibinfo {author} {\bibfnamefont {E.}~\bibnamefont {Soignard}},
  \bibinfo {author} {\bibfnamefont {S.~A.}\ \bibnamefont {Amin}}, \bibinfo
  {author} {\bibfnamefont {M.}~\bibnamefont {Guthrie}}, \bibinfo {author}
  {\bibfnamefont {S.~D.}\ \bibnamefont {Shastri}}, \bibinfo {author}
  {\bibfnamefont {P.~L.}\ \bibnamefont {Lee}}, \ and\ \bibinfo {author}
  {\bibfnamefont {J.~L.}\ \bibnamefont {Yarger}},\ }\bibfield  {title}
  {\enquote {\bibinfo {title} {Structural and topological changes in silica
  glass at pressure},}\ }\href {\doibase 10.1103/PhysRevB.81.054105} {\bibfield
   {journal} {\bibinfo  {journal} {Phys. Rev. B}\ }\textbf {\bibinfo {volume}
  {81}},\ \bibinfo {pages} {054105} (\bibinfo {year} {2010})}\BibitemShut
  {NoStop}%
\bibitem [{\citenamefont {Koziatek}, \citenamefont {Barrat},\ and\
  \citenamefont {Rodney}(2015)}]{Koziatek2015}%
  \BibitemOpen
  \bibfield  {author} {\bibinfo {author} {\bibfnamefont {P.}~\bibnamefont
  {Koziatek}}, \bibinfo {author} {\bibfnamefont {J.}~\bibnamefont {Barrat}}, \
  and\ \bibinfo {author} {\bibfnamefont {D.}~\bibnamefont {Rodney}},\
  }\bibfield  {title} {\enquote {\bibinfo {title} {Short- and medium-range
  orders in as-quenched and deformed $\mathrm{SiO_{2}}$ glasses: An atomistic
  study},}\ }\href {\doibase
  http://dx.doi.org/10.1016/j.jnoncrysol.2015.01.009} {\bibfield  {journal}
  {\bibinfo  {journal} {Journal of Non-Crystalline Solids}\ }\textbf {\bibinfo
  {volume} {414}},\ \bibinfo {pages} {7 -- 15} (\bibinfo {year}
  {2015})}\BibitemShut {NoStop}%
\bibitem [{\citenamefont {Prescher}\ \emph {et~al.}(2017)\citenamefont
  {Prescher}, \citenamefont {Prakapenka}, \citenamefont {Stefanski},
  \citenamefont {Jahn}, \citenamefont {Skinner},\ and\ \citenamefont
  {Wang}}]{Prescher2017}%
  \BibitemOpen
  \bibfield  {author} {\bibinfo {author} {\bibfnamefont {C.}~\bibnamefont
  {Prescher}}, \bibinfo {author} {\bibfnamefont {V.~B.}\ \bibnamefont
  {Prakapenka}}, \bibinfo {author} {\bibfnamefont {J.}~\bibnamefont
  {Stefanski}}, \bibinfo {author} {\bibfnamefont {S.}~\bibnamefont {Jahn}},
  \bibinfo {author} {\bibfnamefont {L.~B.}\ \bibnamefont {Skinner}}, \ and\
  \bibinfo {author} {\bibfnamefont {Y.}~\bibnamefont {Wang}},\ }\bibfield
  {title} {\enquote {\bibinfo {title} {Beyond sixfold coordinated si in sio2
  glass at ultrahigh pressures},}\ }\href {\doibase 10.1073/pnas.1708882114}
  {\bibfield  {journal} {\bibinfo  {journal} {Proceedings of the National
  Academy of Sciences}\ }\textbf {\bibinfo {volume} {114}},\ \bibinfo {pages}
  {10041--10046} (\bibinfo {year} {2017})},\ \Eprint
  {http://arxiv.org/abs/https://www.pnas.org/content/114/38/10041.full.pdf}
  {https://www.pnas.org/content/114/38/10041.full.pdf} \BibitemShut {NoStop}%
\bibitem [{\citenamefont {Murakami}\ \emph {et~al.}(2019)\citenamefont
  {Murakami}, \citenamefont {Kohara}, \citenamefont {Kitamura}, \citenamefont
  {Akola}, \citenamefont {Inoue}, \citenamefont {Hirata}, \citenamefont
  {Hiraoka}, \citenamefont {Onodera}, \citenamefont {Obayashi}, \citenamefont
  {Kalikka}, \citenamefont {Hirao}, \citenamefont {Musso}, \citenamefont
  {Foster}, \citenamefont {Idemoto}, \citenamefont {Sakata},\ and\
  \citenamefont {Ohishi}}]{Murakami2019}%
  \BibitemOpen
  \bibfield  {author} {\bibinfo {author} {\bibfnamefont {M.}~\bibnamefont
  {Murakami}}, \bibinfo {author} {\bibfnamefont {S.}~\bibnamefont {Kohara}},
  \bibinfo {author} {\bibfnamefont {N.}~\bibnamefont {Kitamura}}, \bibinfo
  {author} {\bibfnamefont {J.}~\bibnamefont {Akola}}, \bibinfo {author}
  {\bibfnamefont {H.}~\bibnamefont {Inoue}}, \bibinfo {author} {\bibfnamefont
  {A.}~\bibnamefont {Hirata}}, \bibinfo {author} {\bibfnamefont
  {Y.}~\bibnamefont {Hiraoka}}, \bibinfo {author} {\bibfnamefont
  {Y.}~\bibnamefont {Onodera}}, \bibinfo {author} {\bibfnamefont
  {I.}~\bibnamefont {Obayashi}}, \bibinfo {author} {\bibfnamefont
  {J.}~\bibnamefont {Kalikka}}, \bibinfo {author} {\bibfnamefont
  {N.}~\bibnamefont {Hirao}}, \bibinfo {author} {\bibfnamefont
  {T.}~\bibnamefont {Musso}}, \bibinfo {author} {\bibfnamefont {A.~S.}\
  \bibnamefont {Foster}}, \bibinfo {author} {\bibfnamefont {Y.}~\bibnamefont
  {Idemoto}}, \bibinfo {author} {\bibfnamefont {O.}~\bibnamefont {Sakata}}, \
  and\ \bibinfo {author} {\bibfnamefont {Y.}~\bibnamefont {Ohishi}},\
  }\bibfield  {title} {\enquote {\bibinfo {title} {Ultrahigh-pressure form of
  $\mathrm{Si}{\mathrm{o}}_{2}$ glass with dense pyrite-type crystalline
  homology},}\ }\href {\doibase 10.1103/PhysRevB.99.045153} {\bibfield
  {journal} {\bibinfo  {journal} {Phys. Rev. B}\ }\textbf {\bibinfo {volume}
  {99}},\ \bibinfo {pages} {045153} (\bibinfo {year} {2019})}\BibitemShut
  {NoStop}%
\bibitem [{\citenamefont {Sato}\ and\ \citenamefont
  {Funamori}(2008)}]{Sato2008}%
  \BibitemOpen
  \bibfield  {author} {\bibinfo {author} {\bibfnamefont {T.}~\bibnamefont
  {Sato}}\ and\ \bibinfo {author} {\bibfnamefont {N.}~\bibnamefont
  {Funamori}},\ }\bibfield  {title} {\enquote {\bibinfo {title}
  {Sixfold-coordinated amorphous polymorph of $\mathrm{SiO_{2}}$ under high
  pressure},}\ }\href {\doibase 10.1103/PhysRevLett.101.255502} {\bibfield
  {journal} {\bibinfo  {journal} {Phys. Rev. Lett.}\ }\textbf {\bibinfo
  {volume} {101}},\ \bibinfo {pages} {255502} (\bibinfo {year}
  {2008})}\BibitemShut {NoStop}%
\bibitem [{\citenamefont {Sato}\ and\ \citenamefont
  {Funamori}(2010)}]{Sato2010}%
  \BibitemOpen
  \bibfield  {author} {\bibinfo {author} {\bibfnamefont {T.}~\bibnamefont
  {Sato}}\ and\ \bibinfo {author} {\bibfnamefont {N.}~\bibnamefont
  {Funamori}},\ }\bibfield  {title} {\enquote {\bibinfo {title} {High-pressure
  structural transformation of $\mathrm{SiO_{2}}$ glass up to 100
  $\mathrm{GPa}$},}\ }\href {\doibase 10.1103/PhysRevB.82.184102} {\bibfield
  {journal} {\bibinfo  {journal} {Phys. Rev. B}\ }\textbf {\bibinfo {volume}
  {82}},\ \bibinfo {pages} {184102} (\bibinfo {year} {2010})}\BibitemShut
  {NoStop}%
\bibitem [{\citenamefont {Salmon}\ and\ \citenamefont
  {Zeidler}(2015)}]{Salmon2015}%
  \BibitemOpen
  \bibfield  {author} {\bibinfo {author} {\bibfnamefont {P.~S.}\ \bibnamefont
  {Salmon}}\ and\ \bibinfo {author} {\bibfnamefont {A.}~\bibnamefont
  {Zeidler}},\ }\bibfield  {title} {\enquote {\bibinfo {title} {Networks under
  pressure: the development of in situ high-pressure neutron diffraction for
  glassy and liquid materials},}\ }\href
  {http://stacks.iop.org/0953-8984/27/i=13/a=133201} {\bibfield  {journal}
  {\bibinfo  {journal} {Journal of Physics: Condensed Matter}\ }\textbf
  {\bibinfo {volume} {27}},\ \bibinfo {pages} {133201} (\bibinfo {year}
  {2015})}\BibitemShut {NoStop}%
\bibitem [{\citenamefont {Sonneville}\ \emph {et~al.}(2012)\citenamefont
  {Sonneville}, \citenamefont {Mermet}, \citenamefont {Champagnon},
  \citenamefont {Martinet}, \citenamefont {Margueritat}, \citenamefont
  {de~Ligny}, \citenamefont {Deschamps},\ and\ \citenamefont
  {Balima}}]{Sonneville2012}%
  \BibitemOpen
  \bibfield  {author} {\bibinfo {author} {\bibfnamefont {C.}~\bibnamefont
  {Sonneville}}, \bibinfo {author} {\bibfnamefont {A.}~\bibnamefont {Mermet}},
  \bibinfo {author} {\bibfnamefont {B.}~\bibnamefont {Champagnon}}, \bibinfo
  {author} {\bibfnamefont {C.}~\bibnamefont {Martinet}}, \bibinfo {author}
  {\bibfnamefont {J.}~\bibnamefont {Margueritat}}, \bibinfo {author}
  {\bibfnamefont {D.}~\bibnamefont {de~Ligny}}, \bibinfo {author}
  {\bibfnamefont {T.}~\bibnamefont {Deschamps}}, \ and\ \bibinfo {author}
  {\bibfnamefont {F.}~\bibnamefont {Balima}},\ }\bibfield  {title} {\enquote
  {\bibinfo {title} {Progressive transformations of silica glass upon
  densification},}\ }\href {\doibase 10.1063/1.4754601} {\bibfield  {journal}
  {\bibinfo  {journal} {The Journal of Chemical Physics}\ }\textbf {\bibinfo
  {volume} {137}},\ \bibinfo {pages} {124505} (\bibinfo {year}
  {2012})}\BibitemShut {NoStop}%
\bibitem [{\citenamefont {Sonneville}\ \emph {et~al.}(2013)\citenamefont
  {Sonneville}, \citenamefont {Deschamps}, \citenamefont {Martinet},
  \citenamefont {de~Ligny}, \citenamefont {Mermet},\ and\ \citenamefont
  {Champagnon}}]{Sonneville2013}%
  \BibitemOpen
  \bibfield  {author} {\bibinfo {author} {\bibfnamefont {C.}~\bibnamefont
  {Sonneville}}, \bibinfo {author} {\bibfnamefont {T.}~\bibnamefont
  {Deschamps}}, \bibinfo {author} {\bibfnamefont {C.}~\bibnamefont {Martinet}},
  \bibinfo {author} {\bibfnamefont {D.}~\bibnamefont {de~Ligny}}, \bibinfo
  {author} {\bibfnamefont {A.}~\bibnamefont {Mermet}}, \ and\ \bibinfo {author}
  {\bibfnamefont {B.}~\bibnamefont {Champagnon}},\ }\bibfield  {title}
  {\enquote {\bibinfo {title} {Polyamorphic transitions in silica glass},}\
  }\href {\doibase http://dx.doi.org/10.1016/j.jnoncrysol.2012.12.002}
  {\bibfield  {journal} {\bibinfo  {journal} {Journal of Non-Crystalline
  Solids}\ }\textbf {\bibinfo {volume} {382}},\ \bibinfo {pages} {133 -- 136}
  (\bibinfo {year} {2013})}\BibitemShut {NoStop}%
\bibitem [{\citenamefont {Mackenzie}(1963)}]{MacKenzie1963}%
  \BibitemOpen
  \bibfield  {author} {\bibinfo {author} {\bibfnamefont {J.~D.}\ \bibnamefont
  {Mackenzie}},\ }\bibfield  {title} {\enquote {\bibinfo {title} {High-pressure
  effects on oxide glasses: $\mathrm{II}$, subsequent heat treatment},}\ }\href
  {\doibase 10.1111/j.1151-2916.1963.tb13777.x} {\bibfield  {journal} {\bibinfo
   {journal} {Journal of the American Ceramic Society}\ }\textbf {\bibinfo
  {volume} {46}},\ \bibinfo {pages} {470--476} (\bibinfo {year}
  {1963})}\BibitemShut {NoStop}%
\bibitem [{\citenamefont {Hummel}\ and\ \citenamefont
  {Arndt}(1989)}]{Hummel1989}%
  \BibitemOpen
  \bibfield  {author} {\bibinfo {author} {\bibfnamefont {W.}~\bibnamefont
  {Hummel}}\ and\ \bibinfo {author} {\bibfnamefont {J.}~\bibnamefont {Arndt}},\
  }\bibfield  {title} {\enquote {\bibinfo {title} {Anomalous optical relaxation
  behaviour of densified $\mathrm{SiO_{2}}$ glass},}\ }\href {\doibase
  http://dx.doi.org/10.1016/0022-3093(89)90439-0} {\bibfield  {journal}
  {\bibinfo  {journal} {Journal of Non-Crystalline Solids}\ }\textbf {\bibinfo
  {volume} {109}},\ \bibinfo {pages} {40 -- 46} (\bibinfo {year}
  {1989})}\BibitemShut {NoStop}%
\bibitem [{\citenamefont {Arndt}, \citenamefont {Devine},\ and\ \citenamefont
  {Revesz}(1991)}]{Arndt1991}%
  \BibitemOpen
  \bibfield  {author} {\bibinfo {author} {\bibfnamefont {J.}~\bibnamefont
  {Arndt}}, \bibinfo {author} {\bibfnamefont {R.}~\bibnamefont {Devine}}, \
  and\ \bibinfo {author} {\bibfnamefont {A.}~\bibnamefont {Revesz}},\
  }\bibfield  {title} {\enquote {\bibinfo {title} {Anomalous behaviour of the
  refractive index during the annealing of densified, amorphous
  $\mathrm{SiO_{2}}$},}\ }\href {\doibase
  http://dx.doi.org/10.1016/0022-3093(91)90755-U} {\bibfield  {journal}
  {\bibinfo  {journal} {Journal of Non-Crystalline Solids}\ }\textbf {\bibinfo
  {volume} {131}},\ \bibinfo {pages} {1206 -- 1212} (\bibinfo {year}
  {1991})}\BibitemShut {NoStop}%
\bibitem [{\citenamefont {Surovtsev}\ \emph {et~al.}(2006)\citenamefont
  {Surovtsev}, \citenamefont {Adichtchev}, \citenamefont {Malinovsky},
  \citenamefont {Kalinin},\ and\ \citenamefont {Pal’yanov}}]{Surovtsev2006}%
  \BibitemOpen
  \bibfield  {author} {\bibinfo {author} {\bibfnamefont {N.~V.}\ \bibnamefont
  {Surovtsev}}, \bibinfo {author} {\bibfnamefont {S.~V.}\ \bibnamefont
  {Adichtchev}}, \bibinfo {author} {\bibfnamefont {V.~K.}\ \bibnamefont
  {Malinovsky}}, \bibinfo {author} {\bibfnamefont {A.~A.}\ \bibnamefont
  {Kalinin}}, \ and\ \bibinfo {author} {\bibfnamefont {Y.~N.}\ \bibnamefont
  {Pal’yanov}},\ }\bibfield  {title} {\enquote {\bibinfo {title} {Fast
  relaxation intensity versus silica glass density: existence of sharp
  peculiarity},}\ }\href {http://stacks.iop.org/0953-8984/18/i=19/a=027}
  {\bibfield  {journal} {\bibinfo  {journal} {Journal of Physics: Condensed
  Matter}\ }\textbf {\bibinfo {volume} {18}},\ \bibinfo {pages} {4763}
  (\bibinfo {year} {2006})}\BibitemShut {NoStop}%
\bibitem [{\citenamefont {Martinet}\ \emph {et~al.}(2015)\citenamefont
  {Martinet}, \citenamefont {Kassir-Bodon}, \citenamefont {Deschamps},
  \citenamefont {Cornet}, \citenamefont {Floch}, \citenamefont {Martinez},\
  and\ \citenamefont {Champagnon}}]{Martinet2015}%
  \BibitemOpen
  \bibfield  {author} {\bibinfo {author} {\bibfnamefont {C.}~\bibnamefont
  {Martinet}}, \bibinfo {author} {\bibfnamefont {A.}~\bibnamefont
  {Kassir-Bodon}}, \bibinfo {author} {\bibfnamefont {T.}~\bibnamefont
  {Deschamps}}, \bibinfo {author} {\bibfnamefont {A.}~\bibnamefont {Cornet}},
  \bibinfo {author} {\bibfnamefont {S.~L.}\ \bibnamefont {Floch}}, \bibinfo
  {author} {\bibfnamefont {V.}~\bibnamefont {Martinez}}, \ and\ \bibinfo
  {author} {\bibfnamefont {B.}~\bibnamefont {Champagnon}},\ }\bibfield  {title}
  {\enquote {\bibinfo {title} {Permanently densified $\mathrm{SiO_{2}}$
  glasses: a structural approach},}\ }\href
  {http://stacks.iop.org/0953-8984/27/i=32/a=325401} {\bibfield  {journal}
  {\bibinfo  {journal} {Journal of Physics: Condensed Matter}\ }\textbf
  {\bibinfo {volume} {27}},\ \bibinfo {pages} {325401} (\bibinfo {year}
  {2015})}\BibitemShut {NoStop}%
\bibitem [{\citenamefont {Guerette}\ \emph {et~al.}(2018)\citenamefont
  {Guerette}, \citenamefont {Ackerson}, \citenamefont {Thomas}, \citenamefont
  {Watson},\ and\ \citenamefont {Huang}}]{Guerette2018}%
  \BibitemOpen
  \bibfield  {author} {\bibinfo {author} {\bibfnamefont {M.}~\bibnamefont
  {Guerette}}, \bibinfo {author} {\bibfnamefont {M.~R.}\ \bibnamefont
  {Ackerson}}, \bibinfo {author} {\bibfnamefont {J.}~\bibnamefont {Thomas}},
  \bibinfo {author} {\bibfnamefont {E.~B.}\ \bibnamefont {Watson}}, \ and\
  \bibinfo {author} {\bibfnamefont {L.}~\bibnamefont {Huang}},\ }\bibfield
  {title} {\enquote {\bibinfo {title} {Thermally induced amorphous to amorphous
  transition in hot-compressed silica glass},}\ }\href {\doibase
  10.1063/1.5025592} {\bibfield  {journal} {\bibinfo  {journal} {The Journal of
  Chemical Physics}\ }\textbf {\bibinfo {volume} {148}},\ \bibinfo {pages}
  {194501} (\bibinfo {year} {2018})}\BibitemShut {NoStop}%
\bibitem [{\citenamefont {Cornet}\ \emph {et~al.}(2017)\citenamefont {Cornet},
  \citenamefont {Martinez}, \citenamefont {de~Ligny}, \citenamefont
  {Champagnon},\ and\ \citenamefont {Martinet}}]{Cornet2017}%
  \BibitemOpen
  \bibfield  {author} {\bibinfo {author} {\bibfnamefont {A.}~\bibnamefont
  {Cornet}}, \bibinfo {author} {\bibfnamefont {V.}~\bibnamefont {Martinez}},
  \bibinfo {author} {\bibfnamefont {D.}~\bibnamefont {de~Ligny}}, \bibinfo
  {author} {\bibfnamefont {B.}~\bibnamefont {Champagnon}}, \ and\ \bibinfo
  {author} {\bibfnamefont {C.}~\bibnamefont {Martinet}},\ }\bibfield  {title}
  {\enquote {\bibinfo {title} {Relaxation processes of densified silica
  glass},}\ }\href {\doibase 10.1063/1.4977036} {\bibfield  {journal} {\bibinfo
   {journal} {The Journal of Chemical Physics}\ }\textbf {\bibinfo {volume}
  {146}},\ \bibinfo {pages} {094504} (\bibinfo {year} {2017})}\BibitemShut
  {NoStop}%
\bibitem [{\citenamefont {Doremus}(2002)}]{Doremus2002}%
  \BibitemOpen
  \bibfield  {author} {\bibinfo {author} {\bibfnamefont {R.~H.}\ \bibnamefont
  {Doremus}},\ }\bibfield  {title} {\enquote {\bibinfo {title} {Viscosity of
  silica},}\ }\href {\doibase 10.1063/1.1515132} {\bibfield  {journal}
  {\bibinfo  {journal} {Journal of Applied Physics}\ }\textbf {\bibinfo
  {volume} {92}},\ \bibinfo {pages} {7619--7629} (\bibinfo {year}
  {2002})}\BibitemShut {NoStop}%
\bibitem [{\citenamefont {Arndt}\ and\ \citenamefont
  {Stöffler}(1969)}]{Arndt1969}%
  \BibitemOpen
  \bibfield  {author} {\bibinfo {author} {\bibfnamefont {J.}~\bibnamefont
  {Arndt}}\ and\ \bibinfo {author} {\bibfnamefont {D.}~\bibnamefont
  {Stöffler}},\ }\bibfield  {title} {\enquote {\bibinfo {title} {Anomalous
  chnges in some properties of silica glass densified at very high pressure},}\
  }\href@noop {} {\bibfield  {journal} {\bibinfo  {journal} {Physics and
  Chemistry of Glasses}\ }\textbf {\bibinfo {volume} {10}},\ \bibinfo {pages}
  {117--\&} (\bibinfo {year} {1969})}\BibitemShut {NoStop}%
\bibitem [{\citenamefont {Trease}\ \emph {et~al.}(2017)\citenamefont {Trease},
  \citenamefont {Clark}, \citenamefont {Grandinetti}, \citenamefont
  {Stebbins},\ and\ \citenamefont {Sen}}]{Trease2017}%
  \BibitemOpen
  \bibfield  {author} {\bibinfo {author} {\bibfnamefont {N.~M.}\ \bibnamefont
  {Trease}}, \bibinfo {author} {\bibfnamefont {T.~M.}\ \bibnamefont {Clark}},
  \bibinfo {author} {\bibfnamefont {P.~J.}\ \bibnamefont {Grandinetti}},
  \bibinfo {author} {\bibfnamefont {J.~F.}\ \bibnamefont {Stebbins}}, \ and\
  \bibinfo {author} {\bibfnamefont {S.}~\bibnamefont {Sen}},\ }\bibfield
  {title} {\enquote {\bibinfo {title} {Bond length-bond angle correlation in
  densified silica—results from 17o nmr spectroscopy},}\ }\href {\doibase
  10.1063/1.4983041} {\bibfield  {journal} {\bibinfo  {journal} {The Journal of
  Chemical Physics}\ }\textbf {\bibinfo {volume} {146}},\ \bibinfo {pages}
  {184505} (\bibinfo {year} {2017})}\BibitemShut {NoStop}%
\bibitem [{\citenamefont {Elliott}(1991)}]{Elliott1991a}%
  \BibitemOpen
  \bibfield  {author} {\bibinfo {author} {\bibfnamefont {S.~R.}\ \bibnamefont
  {Elliott}},\ }\bibfield  {title} {\enquote {\bibinfo {title} {Origin of the
  first sharp diffraction peak in the structure factor of covalent glasses},}\
  }\href {\doibase 10.1103/PhysRevLett.67.711} {\bibfield  {journal} {\bibinfo
  {journal} {Phys. Rev. Lett.}\ }\textbf {\bibinfo {volume} {67}},\ \bibinfo
  {pages} {711--714} (\bibinfo {year} {1991})}\BibitemShut {NoStop}%
\bibitem [{\citenamefont {Zaug}, \citenamefont {Soper},\ and\ \citenamefont
  {Clark}(2008)}]{Zaug2008}%
  \BibitemOpen
  \bibfield  {author} {\bibinfo {author} {\bibfnamefont {J.~M.}\ \bibnamefont
  {Zaug}}, \bibinfo {author} {\bibfnamefont {A.~K.}\ \bibnamefont {Soper}}, \
  and\ \bibinfo {author} {\bibfnamefont {S.~M.}\ \bibnamefont {Clark}},\
  }\bibfield  {title} {\enquote {\bibinfo {title} {Pressure-dependent
  structures of amorphous red phosphorus and the origin of the first sharp
  diffraction peaks},}\ }\href {\doibase 10.1038/nmat2290} {\bibfield
  {journal} {\bibinfo  {journal} {Nature Materials}\ }\textbf {\bibinfo
  {volume} {7}},\ \bibinfo {pages} {890--899} (\bibinfo {year}
  {2008})}\BibitemShut {NoStop}%
\bibitem [{\citenamefont {Crupi}\ \emph {et~al.}(2015)\citenamefont {Crupi},
  \citenamefont {Carini}, \citenamefont {Gonz\'alez},\ and\ \citenamefont
  {D'Angelo}}]{Crupi2015}%
  \BibitemOpen
  \bibfield  {author} {\bibinfo {author} {\bibfnamefont {C.}~\bibnamefont
  {Crupi}}, \bibinfo {author} {\bibfnamefont {G.}~\bibnamefont {Carini}},
  \bibinfo {author} {\bibfnamefont {M.}~\bibnamefont {Gonz\'alez}}, \ and\
  \bibinfo {author} {\bibfnamefont {G.}~\bibnamefont {D'Angelo}},\ }\bibfield
  {title} {\enquote {\bibinfo {title} {Origin of the first sharp diffraction
  peak in glasses},}\ }\href {\doibase 10.1103/PhysRevB.92.134206} {\bibfield
  {journal} {\bibinfo  {journal} {Phys. Rev. B}\ }\textbf {\bibinfo {volume}
  {92}},\ \bibinfo {pages} {134206} (\bibinfo {year} {2015})}\BibitemShut
  {NoStop}%
\bibitem [{\citenamefont {Mei}\ \emph {et~al.}(2008)\citenamefont {Mei},
  \citenamefont {Benmore}, \citenamefont {Sen}, \citenamefont {Sharma},\ and\
  \citenamefont {Yarger}}]{Mei2008}%
  \BibitemOpen
  \bibfield  {author} {\bibinfo {author} {\bibfnamefont {Q.}~\bibnamefont
  {Mei}}, \bibinfo {author} {\bibfnamefont {C.~J.}\ \bibnamefont {Benmore}},
  \bibinfo {author} {\bibfnamefont {S.}~\bibnamefont {Sen}}, \bibinfo {author}
  {\bibfnamefont {R.}~\bibnamefont {Sharma}}, \ and\ \bibinfo {author}
  {\bibfnamefont {J.~L.}\ \bibnamefont {Yarger}},\ }\bibfield  {title}
  {\enquote {\bibinfo {title} {Intermediate range order in vitreous silica from
  a partial structure factor analysis},}\ }\href {\doibase
  10.1103/PhysRevB.78.144204} {\bibfield  {journal} {\bibinfo  {journal} {Phys.
  Rev. B}\ }\textbf {\bibinfo {volume} {78}},\ \bibinfo {pages} {144204}
  (\bibinfo {year} {2008})}\BibitemShut {NoStop}%
\bibitem [{\citenamefont {Guinier}\ and\ \citenamefont
  {G.}(1955)}]{Guinier1955}%
  \BibitemOpen
  \bibfield  {author} {\bibinfo {author} {\bibfnamefont {A.}~\bibnamefont
  {Guinier}}\ and\ \bibinfo {author} {\bibfnamefont {F.}~\bibnamefont {G.}},\
  }\href@noop {} {\emph {\bibinfo {title} {Small-Angle Scattering of X-Rays}}}\
  (\bibinfo  {publisher} {John Wiley and Sons},\ \bibinfo {year}
  {1955})\BibitemShut {NoStop}%
\bibitem [{\citenamefont {Levelut}\ and\ \citenamefont
  {Guinier}(1967)}]{Levelut1967}%
  \BibitemOpen
  \bibfield  {author} {\bibinfo {author} {\bibfnamefont {A.~M.}\ \bibnamefont
  {Levelut}}\ and\ \bibinfo {author} {\bibfnamefont {A.}~\bibnamefont
  {Guinier}},\ }\bibfield  {title} {\enquote {\bibinfo {title} {X-rays
  scattering at small angles by homogeneous substances},}\ }\href@noop {}
  {\bibfield  {journal} {\bibinfo  {journal} {Bulletin de la société
  française de minéralogie et de cristallographie}\ }\textbf {\bibinfo
  {volume} {90}},\ \bibinfo {pages} {445--\&} (\bibinfo {year}
  {1967})}\BibitemShut {NoStop}%
\bibitem [{\citenamefont {Levelut}\ \emph {et~al.}(2005)\citenamefont
  {Levelut}, \citenamefont {Faivre}, \citenamefont {Le~Parc}, \citenamefont
  {Champagnon}, \citenamefont {Hazemann},\ and\ \citenamefont
  {Simon}}]{Levelut2005}%
  \BibitemOpen
  \bibfield  {author} {\bibinfo {author} {\bibfnamefont {C.}~\bibnamefont
  {Levelut}}, \bibinfo {author} {\bibfnamefont {A.}~\bibnamefont {Faivre}},
  \bibinfo {author} {\bibfnamefont {R.}~\bibnamefont {Le~Parc}}, \bibinfo
  {author} {\bibfnamefont {B.}~\bibnamefont {Champagnon}}, \bibinfo {author}
  {\bibfnamefont {J.-L.}\ \bibnamefont {Hazemann}}, \ and\ \bibinfo {author}
  {\bibfnamefont {J.-P.}\ \bibnamefont {Simon}},\ }\bibfield  {title} {\enquote
  {\bibinfo {title} {In situ measurements of density fluctuations and
  compressibility in silica glasses as a function of temperature and thermal
  history},}\ }\href {\doibase 10.1103/PhysRevB.72.224201} {\bibfield
  {journal} {\bibinfo  {journal} {Phys. Rev. B}\ }\textbf {\bibinfo {volume}
  {72}},\ \bibinfo {pages} {224201} (\bibinfo {year} {2005})}\BibitemShut
  {NoStop}%
\bibitem [{\citenamefont {Mizuno}, \citenamefont {Mossa},\ and\ \citenamefont
  {Barrat}(2014)}]{Mizuno2014}%
  \BibitemOpen
  \bibfield  {author} {\bibinfo {author} {\bibfnamefont {H.}~\bibnamefont
  {Mizuno}}, \bibinfo {author} {\bibfnamefont {S.}~\bibnamefont {Mossa}}, \
  and\ \bibinfo {author} {\bibfnamefont {J.-L.}\ \bibnamefont {Barrat}},\
  }\bibfield  {title} {\enquote {\bibinfo {title} {Acoustic excitations and
  elastic heterogeneities in disordered solids},}\ }\href {\doibase
  10.1073/pnas.1409490111} {\bibfield  {journal} {\bibinfo  {journal}
  {Proceedings of the National Academy of Sciences}\ }\textbf {\bibinfo
  {volume} {111}},\ \bibinfo {pages} {11949--11954} (\bibinfo {year}
  {2014})}\BibitemShut {NoStop}%
\bibitem [{\citenamefont {Svenson}\ \emph {et~al.}(2017)\citenamefont
  {Svenson}, \citenamefont {Mauro}, \citenamefont {Rzoska}, \citenamefont
  {Bockowski},\ and\ \citenamefont {Smedskjaer}}]{Svenson2017}%
  \BibitemOpen
  \bibfield  {author} {\bibinfo {author} {\bibfnamefont {M.~N.}\ \bibnamefont
  {Svenson}}, \bibinfo {author} {\bibfnamefont {J.~C.}\ \bibnamefont {Mauro}},
  \bibinfo {author} {\bibfnamefont {S.~J.}\ \bibnamefont {Rzoska}}, \bibinfo
  {author} {\bibfnamefont {M.}~\bibnamefont {Bockowski}}, \ and\ \bibinfo
  {author} {\bibfnamefont {M.~M.}\ \bibnamefont {Smedskjaer}},\ }\bibfield
  {title} {\enquote {\bibinfo {title} {Accessing forbidden glass regimes
  through high-pressure sub-tg annealing},}\ }\href {\doibase
  10.1038/srep46631} {\bibfield  {journal} {\bibinfo  {journal} {Scientific
  Reports}\ }\textbf {\bibinfo {volume} {7}} (\bibinfo {year} {2017}),\
  10.1038/srep46631}\BibitemShut {NoStop}%
\bibitem [{\citenamefont {Bauchy}(2012)}]{Bauchy2012}%
  \BibitemOpen
  \bibfield  {author} {\bibinfo {author} {\bibfnamefont {M.}~\bibnamefont
  {Bauchy}},\ }\bibfield  {title} {\enquote {\bibinfo {title} {Structural,
  vibrational, and thermal properties of densified silicates: Insights from
  molecular dynamics},}\ }\href {\doibase 10.1063/1.4738501} {\bibfield
  {journal} {\bibinfo  {journal} {The Journal of Chemical Physics}\ }\textbf
  {\bibinfo {volume} {137}},\ \bibinfo {pages} {044510} (\bibinfo {year}
  {2012})}\BibitemShut {NoStop}%
\bibitem [{\citenamefont {Sokolov}\ \emph {et~al.}(1992)\citenamefont
  {Sokolov}, \citenamefont {Kisliuk}, \citenamefont {Soltwisch},\ and\
  \citenamefont {Quitmann}}]{Sokolov1992}%
  \BibitemOpen
  \bibfield  {author} {\bibinfo {author} {\bibfnamefont {A.~P.}\ \bibnamefont
  {Sokolov}}, \bibinfo {author} {\bibfnamefont {A.}~\bibnamefont {Kisliuk}},
  \bibinfo {author} {\bibfnamefont {M.}~\bibnamefont {Soltwisch}}, \ and\
  \bibinfo {author} {\bibfnamefont {D.}~\bibnamefont {Quitmann}},\ }\bibfield
  {title} {\enquote {\bibinfo {title} {Medium-range order in glasses:
  Comparison of raman and diffraction measurements},}\ }\href {\doibase
  10.1103/PhysRevLett.69.1540} {\bibfield  {journal} {\bibinfo  {journal}
  {Phys. Rev. Lett.}\ }\textbf {\bibinfo {volume} {69}},\ \bibinfo {pages}
  {1540--1543} (\bibinfo {year} {1992})}\BibitemShut {NoStop}%
\bibitem [{\citenamefont {Zanatta}\ \emph {et~al.}(2014)\citenamefont
  {Zanatta}, \citenamefont {Baldi}, \citenamefont {Brusa}, \citenamefont
  {Egger}, \citenamefont {Fontana}, \citenamefont {Gilioli}, \citenamefont
  {Mariazzi}, \citenamefont {Monaco}, \citenamefont {Ravelli},\ and\
  \citenamefont {Sacchetti}}]{Zanatta2014}%
  \BibitemOpen
  \bibfield  {author} {\bibinfo {author} {\bibfnamefont {M.}~\bibnamefont
  {Zanatta}}, \bibinfo {author} {\bibfnamefont {G.}~\bibnamefont {Baldi}},
  \bibinfo {author} {\bibfnamefont {R.~S.}\ \bibnamefont {Brusa}}, \bibinfo
  {author} {\bibfnamefont {W.}~\bibnamefont {Egger}}, \bibinfo {author}
  {\bibfnamefont {A.}~\bibnamefont {Fontana}}, \bibinfo {author} {\bibfnamefont
  {E.}~\bibnamefont {Gilioli}}, \bibinfo {author} {\bibfnamefont
  {S.}~\bibnamefont {Mariazzi}}, \bibinfo {author} {\bibfnamefont
  {G.}~\bibnamefont {Monaco}}, \bibinfo {author} {\bibfnamefont
  {L.}~\bibnamefont {Ravelli}}, \ and\ \bibinfo {author} {\bibfnamefont
  {F.}~\bibnamefont {Sacchetti}},\ }\bibfield  {title} {\enquote {\bibinfo
  {title} {Structural evolution and medium range order in permanently densified
  vitreous $\mathrm{SiO_{2}}$},}\ }\href {\doibase
  10.1103/PhysRevLett.112.045501} {\bibfield  {journal} {\bibinfo  {journal}
  {Phys. Rev. Lett.}\ }\textbf {\bibinfo {volume} {112}},\ \bibinfo {pages}
  {045501} (\bibinfo {year} {2014})}\BibitemShut {NoStop}%
\bibitem [{\citenamefont {Salmon}(2018)}]{Salmon2018}%
  \BibitemOpen
  \bibfield  {author} {\bibinfo {author} {\bibfnamefont {P.~S.}\ \bibnamefont
  {Salmon}},\ }\bibfield  {title} {\enquote {\bibinfo {title} {Chapter 13 -
  densification mechanisms of oxide glasses and melts},}\ }in\ \href {\doibase
  https://doi.org/10.1016/B978-0-12-811301-1.00013-7} {\emph {\bibinfo
  {booktitle} {Magmas Under Pressure}}},\ \bibinfo {editor} {edited by\
  \bibinfo {editor} {\bibfnamefont {Y.}~\bibnamefont {Kono}}\ and\ \bibinfo
  {editor} {\bibfnamefont {C.}~\bibnamefont {Sanloup}}}\ (\bibinfo  {publisher}
  {Elsevier},\ \bibinfo {year} {2018})\ pp.\ \bibinfo {pages} {343 --
  369}\BibitemShut {NoStop}%
\bibitem [{\citenamefont {Guerette}\ \emph {et~al.}(2015)\citenamefont
  {Guerette}, \citenamefont {Ackerson}, \citenamefont {Thomas}, \citenamefont
  {Yuan}, \citenamefont {Watson}, \citenamefont {Walker},\ and\ \citenamefont
  {Huang}}]{Guerette2015}%
  \BibitemOpen
  \bibfield  {author} {\bibinfo {author} {\bibfnamefont {M.}~\bibnamefont
  {Guerette}}, \bibinfo {author} {\bibfnamefont {M.~R.}\ \bibnamefont
  {Ackerson}}, \bibinfo {author} {\bibfnamefont {J.}~\bibnamefont {Thomas}},
  \bibinfo {author} {\bibfnamefont {F.}~\bibnamefont {Yuan}}, \bibinfo {author}
  {\bibfnamefont {E.~B.}\ \bibnamefont {Watson}}, \bibinfo {author}
  {\bibfnamefont {D.}~\bibnamefont {Walker}}, \ and\ \bibinfo {author}
  {\bibfnamefont {L.}~\bibnamefont {Huang}},\ }\bibfield  {title} {\enquote
  {\bibinfo {title} {Structure and properties of silica glass densified in cold
  compression and hot compression},}\ }\href {\doibase 10.1038/srep15343}
  {\bibfield  {journal} {\bibinfo  {journal} {Scientific Reports}\ }\textbf
  {\bibinfo {volume} {5}} (\bibinfo {year} {2015}),\
  10.1038/srep15343}\BibitemShut {NoStop}%
\bibitem [{\citenamefont {Zanatta}\ \emph {et~al.}(2010)\citenamefont
  {Zanatta}, \citenamefont {Baldi}, \citenamefont {Caponi}, \citenamefont
  {Fontana}, \citenamefont {Gilioli}, \citenamefont {Krish}, \citenamefont
  {Masciovecchio}, \citenamefont {Monaco}, \citenamefont {Orsingher},
  \citenamefont {Rossi}, \citenamefont {Ruocco},\ and\ \citenamefont
  {Verbeni}}]{Zanatta2010}%
  \BibitemOpen
  \bibfield  {author} {\bibinfo {author} {\bibfnamefont {M.}~\bibnamefont
  {Zanatta}}, \bibinfo {author} {\bibfnamefont {G.}~\bibnamefont {Baldi}},
  \bibinfo {author} {\bibfnamefont {S.}~\bibnamefont {Caponi}}, \bibinfo
  {author} {\bibfnamefont {A.}~\bibnamefont {Fontana}}, \bibinfo {author}
  {\bibfnamefont {E.}~\bibnamefont {Gilioli}}, \bibinfo {author} {\bibfnamefont
  {M.}~\bibnamefont {Krish}}, \bibinfo {author} {\bibfnamefont
  {C.}~\bibnamefont {Masciovecchio}}, \bibinfo {author} {\bibfnamefont
  {G.}~\bibnamefont {Monaco}}, \bibinfo {author} {\bibfnamefont
  {L.}~\bibnamefont {Orsingher}}, \bibinfo {author} {\bibfnamefont
  {F.}~\bibnamefont {Rossi}}, \bibinfo {author} {\bibfnamefont
  {G.}~\bibnamefont {Ruocco}}, \ and\ \bibinfo {author} {\bibfnamefont
  {R.}~\bibnamefont {Verbeni}},\ }\bibfield  {title} {\enquote {\bibinfo
  {title} {Elastic properties of permanently densified silica: A raman,
  brillouin light, and x-ray scattering study},}\ }\href {\doibase
  10.1103/PhysRevB.81.212201} {\bibfield  {journal} {\bibinfo  {journal} {Phys.
  Rev. B}\ }\textbf {\bibinfo {volume} {81}},\ \bibinfo {pages} {212201}
  (\bibinfo {year} {2010})}\BibitemShut {NoStop}%
\bibitem [{\citenamefont {Inamura}\ \emph {et~al.}(2004)\citenamefont
  {Inamura}, \citenamefont {Katayama}, \citenamefont {Utsumi},\ and\
  \citenamefont {Funakoshi}}]{Inamura2004}%
  \BibitemOpen
  \bibfield  {author} {\bibinfo {author} {\bibfnamefont {Y.}~\bibnamefont
  {Inamura}}, \bibinfo {author} {\bibfnamefont {Y.}~\bibnamefont {Katayama}},
  \bibinfo {author} {\bibfnamefont {W.}~\bibnamefont {Utsumi}}, \ and\ \bibinfo
  {author} {\bibfnamefont {K.-i.}\ \bibnamefont {Funakoshi}},\ }\bibfield
  {title} {\enquote {\bibinfo {title} {Transformations in the
  intermediate-range structure of $\mathrm{SiO_{2}}$ glass under high pressure
  and temperature},}\ }\href {\doibase 10.1103/PhysRevLett.93.015501}
  {\bibfield  {journal} {\bibinfo  {journal} {Phys. Rev. Lett.}\ }\textbf
  {\bibinfo {volume} {93}},\ \bibinfo {pages} {015501} (\bibinfo {year}
  {2004})}\BibitemShut {NoStop}%
\bibitem [{\citenamefont {Shen}\ \emph {et~al.}(2007)\citenamefont {Shen},
  \citenamefont {Liermann}, \citenamefont {Sinogeikin}, \citenamefont {Yang},
  \citenamefont {Hong}, \citenamefont {Yoo},\ and\ \citenamefont
  {Cynn}}]{Shen2007}%
  \BibitemOpen
  \bibfield  {author} {\bibinfo {author} {\bibfnamefont {G.}~\bibnamefont
  {Shen}}, \bibinfo {author} {\bibfnamefont {H.-P.}\ \bibnamefont {Liermann}},
  \bibinfo {author} {\bibfnamefont {S.}~\bibnamefont {Sinogeikin}}, \bibinfo
  {author} {\bibfnamefont {W.}~\bibnamefont {Yang}}, \bibinfo {author}
  {\bibfnamefont {X.}~\bibnamefont {Hong}}, \bibinfo {author} {\bibfnamefont
  {C.-S.}\ \bibnamefont {Yoo}}, \ and\ \bibinfo {author} {\bibfnamefont
  {H.}~\bibnamefont {Cynn}},\ }\bibfield  {title} {\enquote {\bibinfo {title}
  {Distinct thermal behavior of $\mathrm{GeO_{2}}$ glass in tetrahedral,
  intermediate, and octahedral forms},}\ }\href {\doibase
  10.1073/pnas.0703098104} {\bibfield  {journal} {\bibinfo  {journal}
  {Proceedings of the National Academy of Sciences}\ }\textbf {\bibinfo
  {volume} {104}},\ \bibinfo {pages} {14576--14579} (\bibinfo {year} {2007})},\
  \Eprint
  {http://arxiv.org/abs/http://www.pnas.org/content/104/37/14576.full.pdf}
  {http://www.pnas.org/content/104/37/14576.full.pdf} \BibitemShut {NoStop}%
\bibitem [{\citenamefont {Koza}\ \emph {et~al.}(2006)\citenamefont {Koza},
  \citenamefont {Hansen}, \citenamefont {May},\ and\ \citenamefont
  {Schober}}]{Koza2006}%
  \BibitemOpen
  \bibfield  {author} {\bibinfo {author} {\bibfnamefont {M.}~\bibnamefont
  {Koza}}, \bibinfo {author} {\bibfnamefont {T.}~\bibnamefont {Hansen}},
  \bibinfo {author} {\bibfnamefont {R.}~\bibnamefont {May}}, \ and\ \bibinfo
  {author} {\bibfnamefont {H.}~\bibnamefont {Schober}},\ }\bibfield  {title}
  {\enquote {\bibinfo {title} {Link between the diversity, heterogeneity and
  kinetic properties of amorphous ice structures},}\ }\href {\doibase
  https://doi.org/10.1016/j.jnoncrysol.2006.02.162} {\bibfield  {journal}
  {\bibinfo  {journal} {Journal of Non-Crystalline Solids}\ }\textbf {\bibinfo
  {volume} {352}},\ \bibinfo {pages} {4988 -- 4993} (\bibinfo {year} {2006})},\
  \bibinfo {note} {proceedings of the 5th International Discussion Meeting on
  Relaxations in Complex Systems}\BibitemShut {NoStop}%
\bibitem [{\citenamefont {Koza}, \citenamefont {May},\ and\ \citenamefont
  {Schober}(2007)}]{Koza2007}%
  \BibitemOpen
  \bibfield  {author} {\bibinfo {author} {\bibfnamefont {M.~M.}\ \bibnamefont
  {Koza}}, \bibinfo {author} {\bibfnamefont {R.~P.}\ \bibnamefont {May}}, \
  and\ \bibinfo {author} {\bibfnamefont {H.}~\bibnamefont {Schober}},\
  }\bibfield  {title} {\enquote {\bibinfo {title} {{On the heterogeneous
  character of water's amorphous polymorphism}},}\ }\href {\doibase
  10.1107/S0021889807004992} {\bibfield  {journal} {\bibinfo  {journal}
  {Journal of Applied Crystallography}\ }\textbf {\bibinfo {volume} {40}},\
  \bibinfo {pages} {s517--s521} (\bibinfo {year} {2007})}\BibitemShut {NoStop}%
\bibitem [{\citenamefont {Kim}\ \emph {et~al.}(2017)\citenamefont {Kim},
  \citenamefont {Sp{\"a}h}, \citenamefont {Pathak}, \citenamefont {Perakis},
  \citenamefont {Mariedahl}, \citenamefont {Amann-Winkel}, \citenamefont
  {Sellberg}, \citenamefont {Lee}, \citenamefont {Kim}, \citenamefont {Park},
  \citenamefont {Nam}, \citenamefont {Katayama},\ and\ \citenamefont
  {Nilsson}}]{Kim2017}%
  \BibitemOpen
  \bibfield  {author} {\bibinfo {author} {\bibfnamefont {K.~H.}\ \bibnamefont
  {Kim}}, \bibinfo {author} {\bibfnamefont {A.}~\bibnamefont {Sp{\"a}h}},
  \bibinfo {author} {\bibfnamefont {H.}~\bibnamefont {Pathak}}, \bibinfo
  {author} {\bibfnamefont {F.}~\bibnamefont {Perakis}}, \bibinfo {author}
  {\bibfnamefont {D.}~\bibnamefont {Mariedahl}}, \bibinfo {author}
  {\bibfnamefont {K.}~\bibnamefont {Amann-Winkel}}, \bibinfo {author}
  {\bibfnamefont {J.~A.}\ \bibnamefont {Sellberg}}, \bibinfo {author}
  {\bibfnamefont {J.~H.}\ \bibnamefont {Lee}}, \bibinfo {author} {\bibfnamefont
  {S.}~\bibnamefont {Kim}}, \bibinfo {author} {\bibfnamefont {J.}~\bibnamefont
  {Park}}, \bibinfo {author} {\bibfnamefont {K.~H.}\ \bibnamefont {Nam}},
  \bibinfo {author} {\bibfnamefont {T.}~\bibnamefont {Katayama}}, \ and\
  \bibinfo {author} {\bibfnamefont {A.}~\bibnamefont {Nilsson}},\ }\bibfield
  {title} {\enquote {\bibinfo {title} {Maxima in the thermodynamic response and
  correlation functions of deeply supercooled water},}\ }\href {\doibase
  10.1126/science.aap8269} {\bibfield  {journal} {\bibinfo  {journal}
  {Science}\ }\textbf {\bibinfo {volume} {358}},\ \bibinfo {pages} {1589--1593}
  (\bibinfo {year} {2017})},\ \Eprint
  {http://arxiv.org/abs/https://science.sciencemag.org/content/358/6370/1589.full.pdf}
  {https://science.sciencemag.org/content/358/6370/1589.full.pdf} \BibitemShut
  {NoStop}%
\end{thebibliography}%


\providecommand{\noopsort}[1]{}\providecommand{\singleletter}[1]{#1}%
\begin{thebibliography}{8}%
\makeatletter
\providecommand \@ifxundefined [1]{%
 \@ifx{#1\undefined}
}%
\providecommand \@ifnum [1]{%
 \ifnum #1\expandafter \@firstoftwo
 \else \expandafter \@secondoftwo
 \fi
}%
\providecommand \@ifx [1]{%
 \ifx #1\expandafter \@firstoftwo
 \else \expandafter \@secondoftwo
 \fi
}%
\providecommand \natexlab [1]{#1}%
\providecommand \enquote  [1]{``#1''}%
\providecommand \bibnamefont  [1]{#1}%
\providecommand \bibfnamefont [1]{#1}%
\providecommand \citenamefont [1]{#1}%
\providecommand \href@noop [0]{\@secondoftwo}%
\providecommand \href [0]{\begingroup \@sanitize@url \@href}%
\providecommand \@href[1]{\@@startlink{#1}\@@href}%
\providecommand \@@href[1]{\endgroup#1\@@endlink}%
\providecommand \@sanitize@url [0]{\catcode `\\12\catcode `\$12\catcode
  `\&12\catcode `\#12\catcode `\^12\catcode `\_12\catcode `\%12\relax}%
\providecommand \@@startlink[1]{}%
\providecommand \@@endlink[0]{}%
\providecommand \url  [0]{\begingroup\@sanitize@url \@url }%
\providecommand \@url [1]{\endgroup\@href {#1}{\urlprefix }}%
\providecommand \urlprefix  [0]{URL }%
\providecommand \Eprint [0]{\href }%
\providecommand \doibase [0]{http://dx.doi.org/}%
\providecommand \selectlanguage [0]{\@gobble}%
\providecommand \bibinfo  [0]{\@secondoftwo}%
\providecommand \bibfield  [0]{\@secondoftwo}%
\providecommand \translation [1]{[#1]}%
\providecommand \BibitemOpen [0]{}%
\providecommand \bibitemStop [0]{}%
\providecommand \bibitemNoStop [0]{.\EOS\space}%
\providecommand \EOS [0]{\spacefactor3000\relax}%
\providecommand \BibitemShut  [1]{\csname bibitem#1\endcsname}%
\let\auto@bib@innerbib\@empty
\bibitem [{\citenamefont {Levelut}\ \emph {et~al.}(2005)\citenamefont
  {Levelut}, \citenamefont {Faivre}, \citenamefont {Le~Parc}, \citenamefont
  {Champagnon}, \citenamefont {Hazemann},\ and\ \citenamefont
  {Simon}}]{Levelut2005}%
  \BibitemOpen
  \bibfield  {author} {\bibinfo {author} {\bibfnamefont {C.}~\bibnamefont
  {Levelut}}, \bibinfo {author} {\bibfnamefont {A.}~\bibnamefont {Faivre}},
  \bibinfo {author} {\bibfnamefont {R.}~\bibnamefont {Le~Parc}}, \bibinfo
  {author} {\bibfnamefont {B.}~\bibnamefont {Champagnon}}, \bibinfo {author}
  {\bibfnamefont {J.-L.}\ \bibnamefont {Hazemann}}, \ and\ \bibinfo {author}
  {\bibfnamefont {J.-P.}\ \bibnamefont {Simon}},\ }\bibfield  {title} {\enquote
  {\bibinfo {title} {In situ measurements of density fluctuations and
  compressibility in silica glasses as a function of temperature and thermal
  history},}\ }\href {\doibase 10.1103/PhysRevB.72.224201} {\bibfield
  {journal} {\bibinfo  {journal} {Phys. Rev. B}\ }\textbf {\bibinfo {volume}
  {72}},\ \bibinfo {pages} {224201} (\bibinfo {year} {2005})}\BibitemShut
  {NoStop}%
\bibitem [{\citenamefont {Guinier}\ and\ \citenamefont
  {G.}(1955)}]{Guinier1955}%
  \BibitemOpen
  \bibfield  {author} {\bibinfo {author} {\bibfnamefont {A.}~\bibnamefont
  {Guinier}}\ and\ \bibinfo {author} {\bibfnamefont {F.}~\bibnamefont {G.}},\
  }\href@noop {} {\emph {\bibinfo {title} {Small-Angle Scattering of X-Rays}}}\
  (\bibinfo  {publisher} {John Wiley and Sons},\ \bibinfo {year}
  {1955})\BibitemShut {NoStop}%
\bibitem [{\citenamefont {Levelut}\ and\ \citenamefont
  {Guinier}(1967)}]{Levelut1967}%
  \BibitemOpen
  \bibfield  {author} {\bibinfo {author} {\bibfnamefont {A.~M.}\ \bibnamefont
  {Levelut}}\ and\ \bibinfo {author} {\bibfnamefont {A.}~\bibnamefont
  {Guinier}},\ }\bibfield  {title} {\enquote {\bibinfo {title} {X-rays
  scattering at small angles by homogeneous substances},}\ }\href@noop {}
  {\bibfield  {journal} {\bibinfo  {journal} {Bulletin de la société
  française de minéralogie et de cristallographie}\ }\textbf {\bibinfo
  {volume} {90}},\ \bibinfo {pages} {445--\&} (\bibinfo {year}
  {1967})}\BibitemShut {NoStop}%
\bibitem [{\citenamefont {Susman}\ \emph {et~al.}(1991)\citenamefont {Susman},
  \citenamefont {Volin}, \citenamefont {Montague},\ and\ \citenamefont
  {Price}}]{Susman1991}%
  \BibitemOpen
  \bibfield  {author} {\bibinfo {author} {\bibfnamefont {S.}~\bibnamefont
  {Susman}}, \bibinfo {author} {\bibfnamefont {K.~J.}\ \bibnamefont {Volin}},
  \bibinfo {author} {\bibfnamefont {D.~G.}\ \bibnamefont {Montague}}, \ and\
  \bibinfo {author} {\bibfnamefont {D.~L.}\ \bibnamefont {Price}},\ }\bibfield
  {title} {\enquote {\bibinfo {title} {Temperature dependence of the first
  sharp diffraction peak in vitreous silica},}\ }\href {\doibase
  10.1103/PhysRevB.43.11076} {\bibfield  {journal} {\bibinfo  {journal} {Phys.
  Rev. B}\ }\textbf {\bibinfo {volume} {43}},\ \bibinfo {pages} {11076--11081}
  (\bibinfo {year} {1991})}\BibitemShut {NoStop}%
\bibitem [{\citenamefont {Meade}, \citenamefont {Hemley},\ and\ \citenamefont
  {Mao}(1992)}]{Meade1992}%
  \BibitemOpen
  \bibfield  {author} {\bibinfo {author} {\bibfnamefont {C.}~\bibnamefont
  {Meade}}, \bibinfo {author} {\bibfnamefont {R.~J.}\ \bibnamefont {Hemley}}, \
  and\ \bibinfo {author} {\bibfnamefont {H.~K.}\ \bibnamefont {Mao}},\
  }\bibfield  {title} {\enquote {\bibinfo {title} {High-pressure
  $\mathrm{X}$-ray diffraction of $\mathrm{SiO_{2}}$ glass},}\ }\href {\doibase
  10.1103/PhysRevLett.69.1387} {\bibfield  {journal} {\bibinfo  {journal}
  {Phys. Rev. Lett.}\ }\textbf {\bibinfo {volume} {69}},\ \bibinfo {pages}
  {1387--1390} (\bibinfo {year} {1992})}\BibitemShut {NoStop}%
\bibitem [{\citenamefont {Sato}\ and\ \citenamefont
  {Funamori}(2008)}]{Sato2008}%
  \BibitemOpen
  \bibfield  {author} {\bibinfo {author} {\bibfnamefont {T.}~\bibnamefont
  {Sato}}\ and\ \bibinfo {author} {\bibfnamefont {N.}~\bibnamefont
  {Funamori}},\ }\bibfield  {title} {\enquote {\bibinfo {title}
  {Sixfold-coordinated amorphous polymorph of $\mathrm{SiO_{2}}$ under high
  pressure},}\ }\href {\doibase 10.1103/PhysRevLett.101.255502} {\bibfield
  {journal} {\bibinfo  {journal} {Phys. Rev. Lett.}\ }\textbf {\bibinfo
  {volume} {101}},\ \bibinfo {pages} {255502} (\bibinfo {year}
  {2008})}\BibitemShut {NoStop}%
\bibitem [{\citenamefont {Cornet}\ \emph {et~al.}(2017)\citenamefont {Cornet},
  \citenamefont {Martinez}, \citenamefont {de~Ligny}, \citenamefont
  {Champagnon},\ and\ \citenamefont {Martinet}}]{Cornet2017}%
  \BibitemOpen
  \bibfield  {author} {\bibinfo {author} {\bibfnamefont {A.}~\bibnamefont
  {Cornet}}, \bibinfo {author} {\bibfnamefont {V.}~\bibnamefont {Martinez}},
  \bibinfo {author} {\bibfnamefont {D.}~\bibnamefont {de~Ligny}}, \bibinfo
  {author} {\bibfnamefont {B.}~\bibnamefont {Champagnon}}, \ and\ \bibinfo
  {author} {\bibfnamefont {C.}~\bibnamefont {Martinet}},\ }\bibfield  {title}
  {\enquote {\bibinfo {title} {Relaxation processes of densified silica
  glass},}\ }\href {\doibase 10.1063/1.4977036} {\bibfield  {journal} {\bibinfo
   {journal} {The Journal of Chemical Physics}\ }\textbf {\bibinfo {volume}
  {146}},\ \bibinfo {pages} {094504} (\bibinfo {year} {2017})}\BibitemShut
  {NoStop}%
\bibitem [{\citenamefont {Brückner}(1970)}]{Bruckner1970}%
  \BibitemOpen
  \bibfield  {author} {\bibinfo {author} {\bibfnamefont {R.}~\bibnamefont
  {Brückner}},\ }\bibfield  {title} {\enquote {\bibinfo {title} {Properties
  and structure of vitreous silica. i},}\ }\href {\doibase
  https://doi.org/10.1016/0022-3093(70)90190-0} {\bibfield  {journal} {\bibinfo
   {journal} {Journal of Non-Crystalline Solids}\ }\textbf {\bibinfo {volume}
  {5}},\ \bibinfo {pages} {123 -- 175} (\bibinfo {year} {1970})}\BibitemShut
  {NoStop}%
\end{thebibliography}%

\end{document}


\preprint{AIP/123-QED}

\title{Evidence of polyamorphic transitions during densified $\mathrm{SiO_{2}}$ glass annealing}

\author{Antoine Cornet}

 \altaffiliation{Now at the University of Manchester: antoine.cornet@manchester.ac.uk}
\affiliation{ Institut Lumi\`{e}re Mati\`{e}re, Univ Lyon, Universit\'{e} Claude Bernard Lyon 1, CNRS, F-69622 Villeurbanne, France
}%
\author{Christine Martinet}%

\affiliation{ Institut Lumi\`{e}re Mati\`{e}re, Univ Lyon, Universit\'{e} Claude Bernard Lyon 1, CNRS, F-69622 Villeurbanne, France
}%
\author{Val\'{e}rie Martinez}%

\affiliation{ Institut Lumi\`{e}re Mati\`{e}re, Univ Lyon, Universit\'{e} Claude Bernard Lyon 1, CNRS, F-69622 Villeurbanne, France
}%

\author{Dominique de Ligny}

\affiliation{
Department  of Materials Science, Glass and Ceramics, University Erlangen-N\"{u}rnberg, Martensstra., D-91058 Erlangen, Germany
}%

\date{\today}
             
\maketitle

\section{Extraction and treatment of SAXS and WAXS data\\}

\paragraph{Normalization\\}

For each experimental configuration, i.e. for each change in the ensemble furnace-lid-windows, a run was completed without any sample over the entire temperature range. The spectra corresponding with SAXS and WAXS data provide the background to normalize the experimental data at the correct temperature. After dark field correction, each experimental spectrum is normalized using Levelut's procedure\cite{Levelut2005}:
\begin{flushleft}
\begin{equation}
\begin{split}
I \left( q \right) \propto \left( t \ \mathrm{ln}\left( \frac{1}{t} \right) \right) ^{-1} \times \ \ \ \ \ \ \ \ \ \ \ \ \ \ \ \ \ \ \ \ \ \ \ \ \ \ \ \ \ \ \ \ \ \\ \left[ I^{S+B} \left( q \right) / m_{1} ^{S+B} - I^{B} \left( q \right) / m_{1} ^{B} \right] \times \frac{m_{1} ^{S+B}}{m_{0} ^{S+B}}
\end{split}
\end{equation}
\end{flushleft}
\begin{flushleft}
\begin{equation}
	t = \dfrac{m_{1} ^{S+B} \times m_{0} ^{B}}{m_{0} ^{S+B} \times m_{1} ^{B}} \ \ \ \ \ \ \ \ \ \ \ \ \ \ \ \ \ \ \ \ \ \ \ \ \ \ \ \ \ \ \ \ \ \ \ \ \ \ \ \ \ \ \ \ \ \ \ \
\end{equation}
\end{flushleft}

In these expressions, $m_{0}$ and $m_{1}$ correspond to the flux of the beam respectively before and after the furnace, $t$ corresponds to the transmission and the suffixes $^{S}$ and $^{B}$ refer to sample and background. The spectra are then corrected from fluctuations of the beam intensity, and normalized to the absorption of the sample.\\

\paragraph{SAXS data\\}
\begin{figure}
\center
\includegraphics[width=9cm]{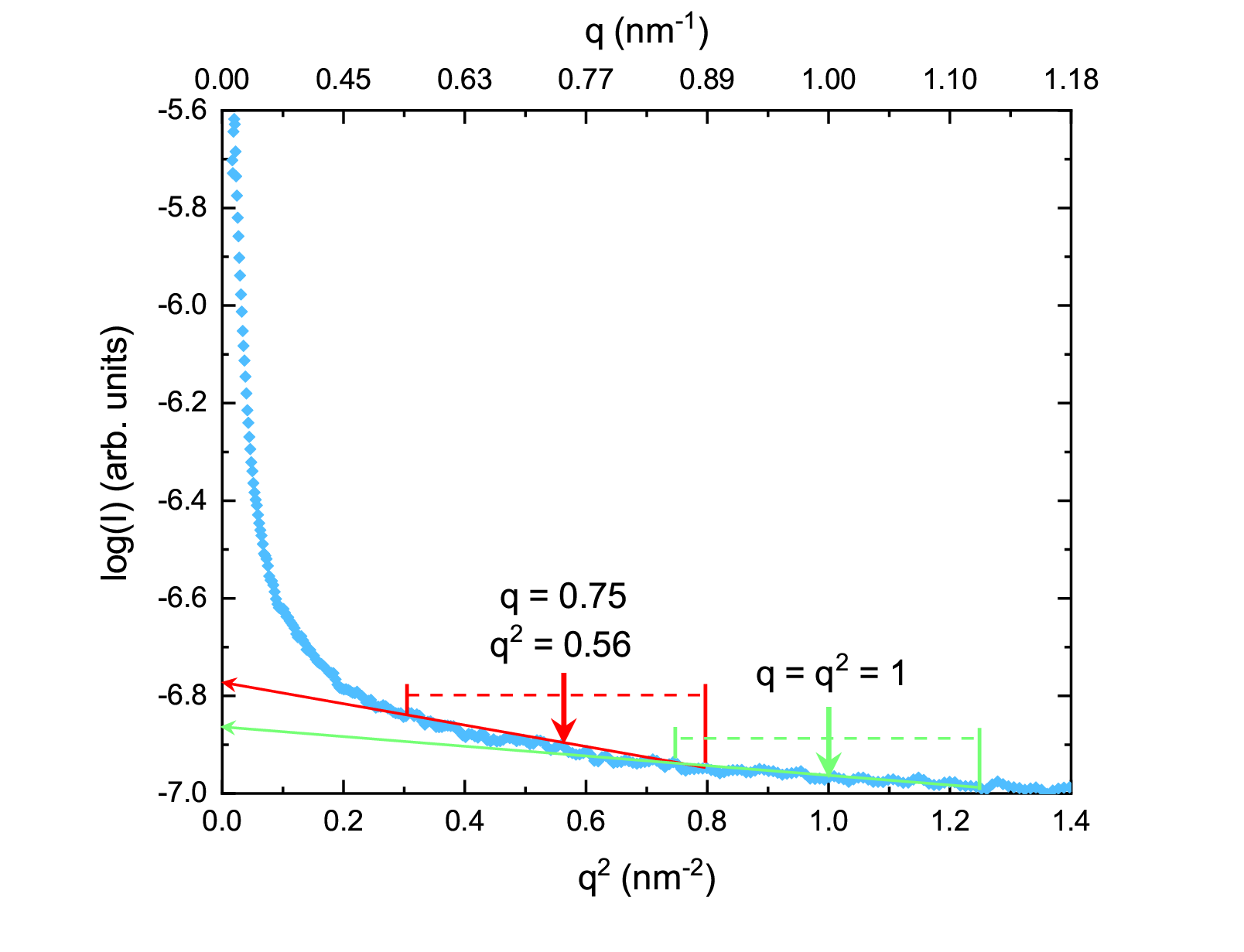}
\caption{SAXS intensity as a function of q or $\mathrm{q^{2}}$, focus on the low q region.}
\label{fig:extraction_I0}
\end{figure}
\begin{figure}
\center
\includegraphics[width=8cm]{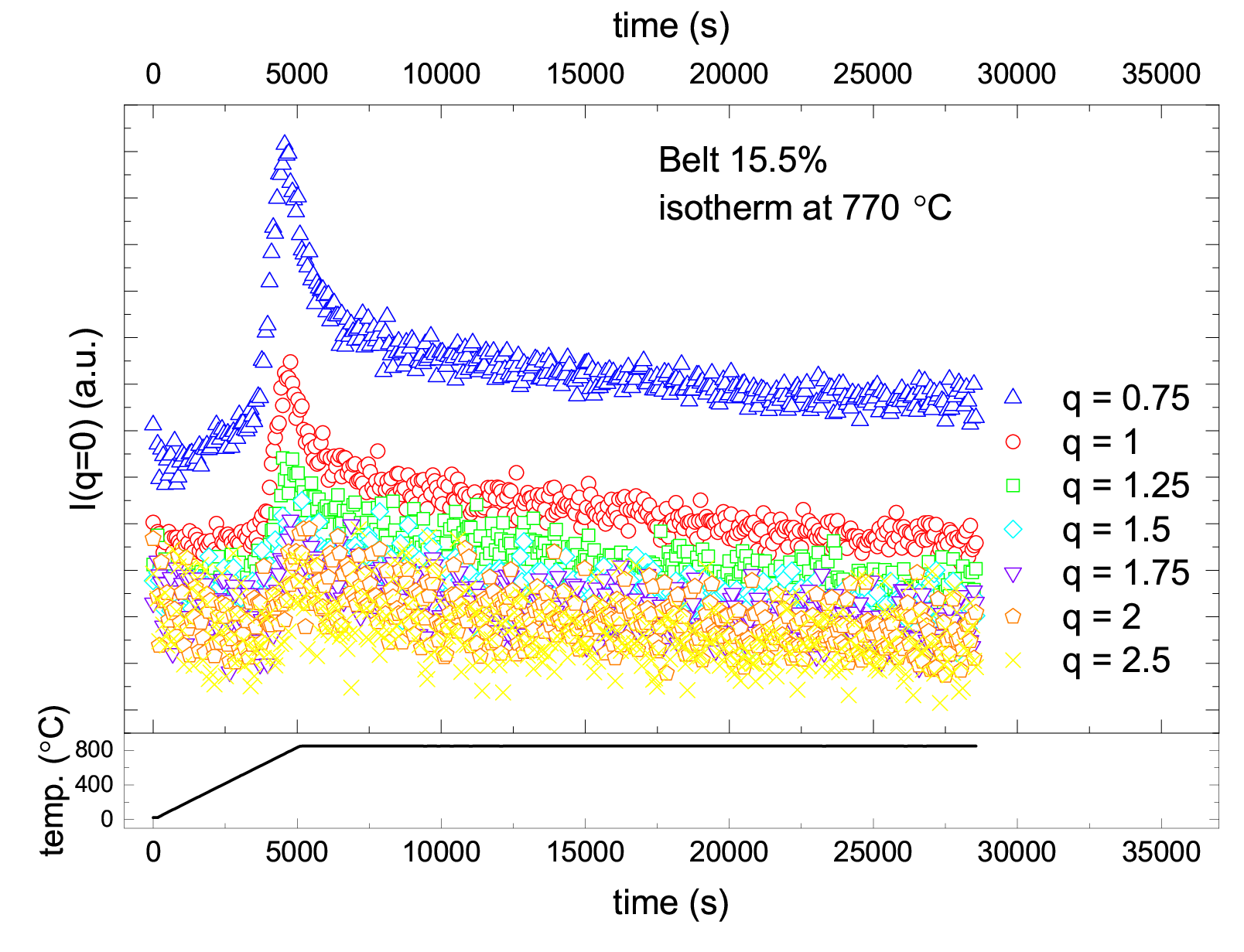}
\caption{Evolution of the intensity extrapolated at q=0 for extrapolation around different values of $\mathrm{q_{nominal}}$: 0.75, 1, 1.25, 1.5, 1.75, 2 and 2.5 $\mathrm{nm^{-1}}$. The data from the entire experiment are represented as a function of time in the upper panel, while the evolution of temperature is represented in the lower panel.}
\label{fig:I0_q_D750R850}
\end{figure}
\begin{figure}[h!]
\center
\includegraphics[width=8cm]{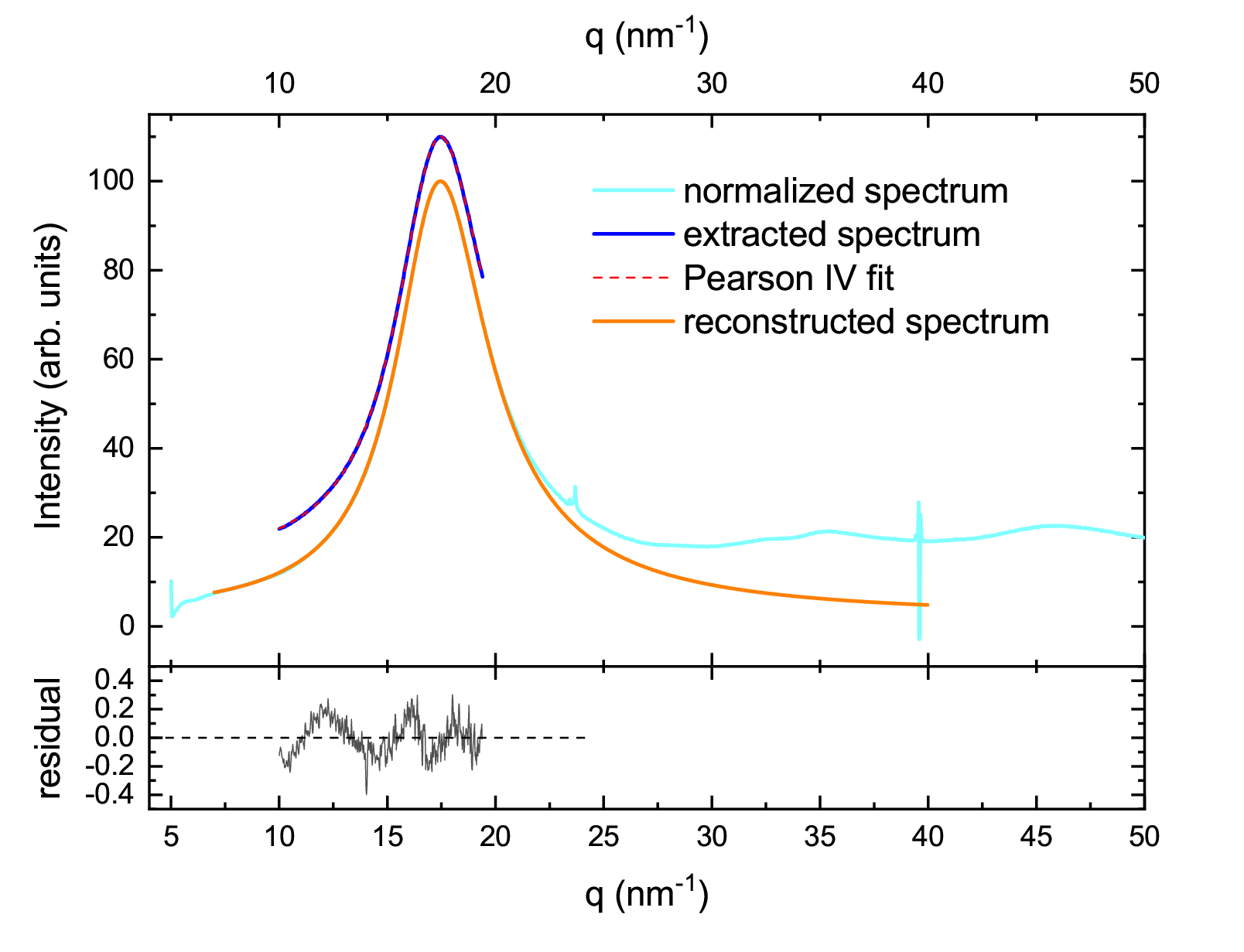}
\caption{Characteristic diffraction pattern of silica showing the first sharp diffraction peak, fit and reconstructed FSDP in the upper panel. In the lower panel, the residuals of the fit are shown.}
\label{fig:extraction_WAXS}
\end{figure}
\begin{figure*}
\center
\includegraphics[width=17cm]{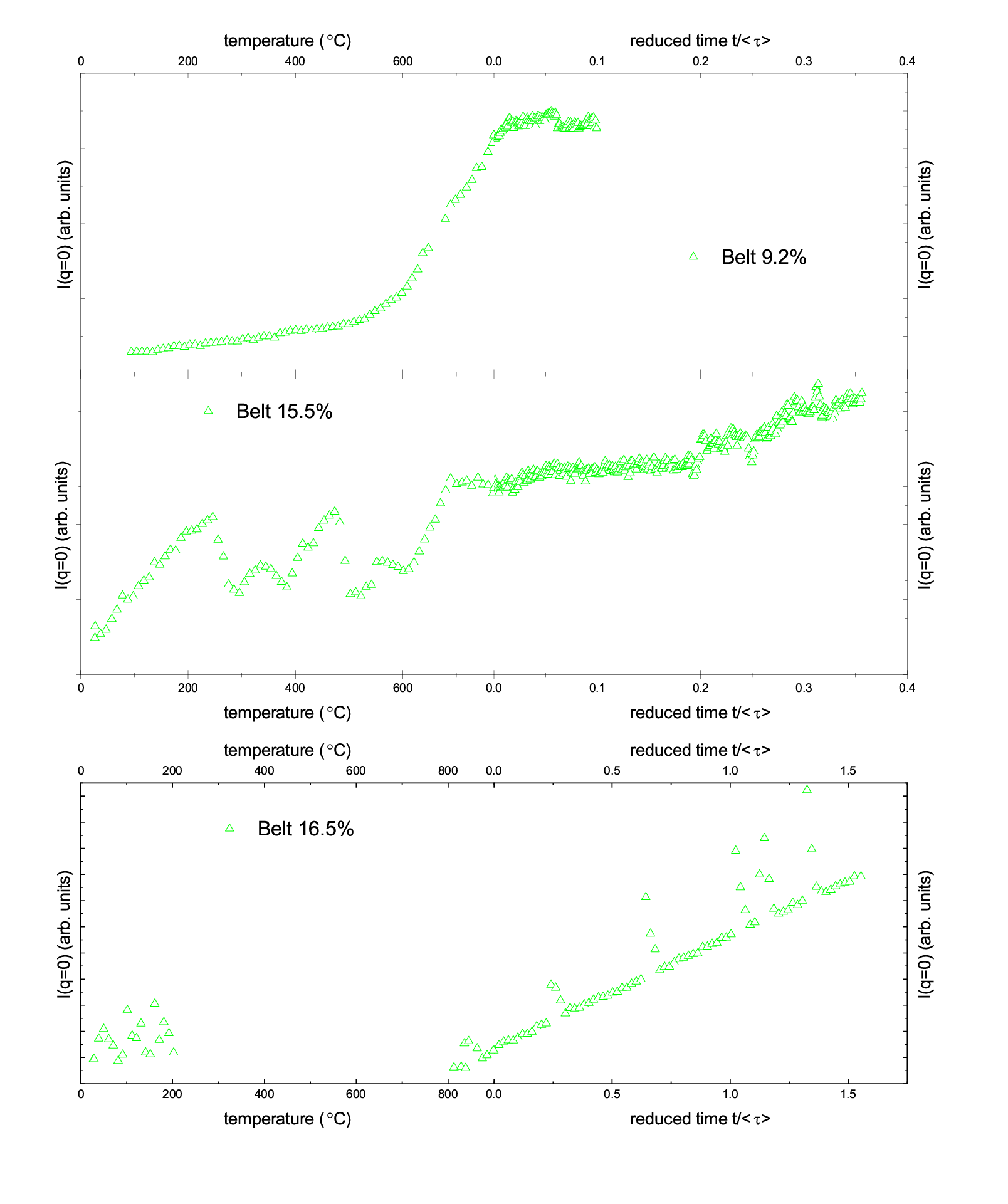}
\caption{Evolution of the intensity extrapolated at q=0 for extrapolation made by the use of a linear fit of the raw data between $\mathrm{q^{2} = 0.02 \ and \ q^{2} = 0.06 \ nm^{-1}}$ for the three different samples. The horizontal axis represent both the temperature and the annealing time.}
\label{fig:low_q}
\end{figure*}

According to Guinier,\cite{Guinier1955, Levelut1967} to extract the intensity at $q = 0$ we have to perform a linear fit of the data in the $ \left[ q^{2};\mathrm{ln} \left( I \right) \right] $ space. To perform the linear extrapolation, we considered several values of q ($\mathrm{q_{nominal} = 0.75, 1, 1.25, 1.75, ..., 4 \ nm^{-1}}$), which we elevated at the power 2. We selected the two points in the experimental spectrum at an equal distance $\mathrm{\Delta q ^{2} = 0.25 nm^{-2}}$ of the nominal $\mathrm{q^{2}}$ values. These two points form a line, that we extrapolate to the value $\mathrm{I \left( q = 0 \right) }$. This is shown in figure \ref{fig:extraction_I0} for a random spectrum, for the two first values of $\mathrm{q_{nominal}}$. Extraction from two points only might not be sufficient compared to a proper fit of a serie of points within an interval, but the result is the same (within uncertainties) regardless the extrapolation technic. This is due to the choice of the value of $\mathrm{\Delta q^{2}}$. The results for some selected values of $\mathrm{q_{nominal}}$ for the experiment conducted on the Belt 15.5\% sample, with a steady temperature of 770\degree C, are shown in figure \ref{fig:I0_q_D750R850}. For an extrapolation made using low values of q, namely $\mathrm{q_{nominal} = 0.75 \ nm^{-1}}$ and $\mathrm{q_{nominal} = 1 \ nm^{-1}}$, the maximum is clearly visible, with a reasonable signal over noise ratio. As the nominal value of q increases, the rise of intensity leading to the maximum decreases, down to the point where it is comparable to the noise. Given that observation, we used the parameters $\mathrm{q_{nominal} = 0.75 \ nm}$ and $\mathrm{\Delta q ^{2} = 0.25 \ nm^{-2}}$ for all the extraction of $\mathrm{I \left( q=0 \right) }$ in the present study.

On the figure \ref{fig:extraction_I0}, one sees that the extrapolation could be done at lower q, because the slope of the curve change dramatically. However, in this case and with our experimental set-up, we reach the limit of the detection window. The extrapolation results from this very low q region for the three densified samples are presented in figure \ref{fig:low_q}, with a continuity between the heating ramp and the isothermal annealings in the horizontal axis. We can see, especially in the heating part of the Belt 15.5\% sample, several maxima, due to several drops in $\mathrm{I \left( q=0 \right) }$. This more than likely due to erratic data at the edge of the detection window. The almost perfectly linear increase for the Belt 16.5\% is questionnable as well. Thus, we did not extend our study to values of q ($\mathrm{- \sqrt{\Delta q ^{2}}}$) lower than $\mathrm{0.75 \ nm^{-1}}$.\\

\paragraph{WAXS data\\}

As highlighted in the main text of this paper, the detection window was fixed because of the experimental set-up. Expanding this detection window in the low-q region would have been done at the expense of considerably cutting the SAXS signal. Thus, the low q tail of the First Sharp Diffraction Peak (FSDP) could not be recorded, and the minimum of the FSDP could not be determined satisfactorily. This implies troubles in the determination of the half-maximum, and finally in $\mathrm{\Delta L_{FSDP}}$.

Because of this, all the work done on the WAXS data was perfomed by fitting a Pearson IV distribution to the FSDP. This function, whose expression is shown below, was derived to model asymmetric probability distribution, which makes it suitable for fitting the FSDP. The expression of the function is:
\begin{equation}
\begin{split}
I \left( q \right) = k \left[ 1 + \left( \frac{q - \lambda}{\alpha} \right) ^{2} \right] ^{-m} \\
\times \mathrm{exp} \left[ - \nu \mathrm{tan^{-1}} \left( \frac{q - \lambda}{\alpha} \right) \right]
\end{split} 
\end{equation}
with the normalizing factor being equal to:
\begin{equation}
k = \dfrac{2^{2m-2} \vert \Gamma \left( m + i \nu / 2 \right)  \vert ^{2}}{\pi \alpha \Gamma \left(2 m - 1 \right)}
\end{equation}
where $\alpha > 0$ and $m > 1/2$ are scale and shape parameters, $\lambda$ a location parameter and $\nu$ an additional parameter. $\Gamma$ is the gamma function.  
This function admits a mode at q equals:
\begin{equation}
q_{maximum} = \lambda - \frac{a \nu}{2 m}
\end{equation}
The different steps of the fitting process are highlighted in figure \ref{fig:extraction_WAXS}.
The normalized cyan (color online) spectrum, between 5 and 50 $\mathrm{nm^{-1}}$, is a random spectrum acquired during the heating ramp in the Belt 16.5\% experiment. A portion of this spectrum, slightly shifted vertically, is considered for the fit. This portion is as wide as possible in the low q region, but limited in the high q region, with the limit fixed at $\mathrm{q = q_{maximum} + 2 \ nm^{-1}}$. This is to avoid significant contribution of the "classical" structure factor (the one originating from the tetrahedral structure of the glass). The fit is performed on that portion of the spectrum, and visible here in a dashed red line (color online). The residual of this fit is shown in the lower part of the figure, and can be seen to be between two and three orders of magnitude lower than the signal. Thus the fit can be considered as good. Moreover, this fit quality is representative of all the treated data. 

Using the parameters given by the fit (performed with Origin software using the Levenberg Marquardt algorithm), we were able to reconstruct the FSDP in a wider q region. This corresponds to the orange spectrum (color online), which spreads from 7 to 40 $\mathrm{nm^{-1}}$ in figure \ref{fig:extraction_WAXS}. We used the value of the reconstructed FSDP at $\mathrm{q = 40 nm^{-1}}$ as the minimum. We were then able to derive $\mathrm{\Delta L_{FSDP}}$, which is defined as the full width at half maximum.

\section{WAXS data of pristine silica and estimation of uncertainties\\}

\begin{figure}[h!]
\center
\includegraphics[width=8cm]{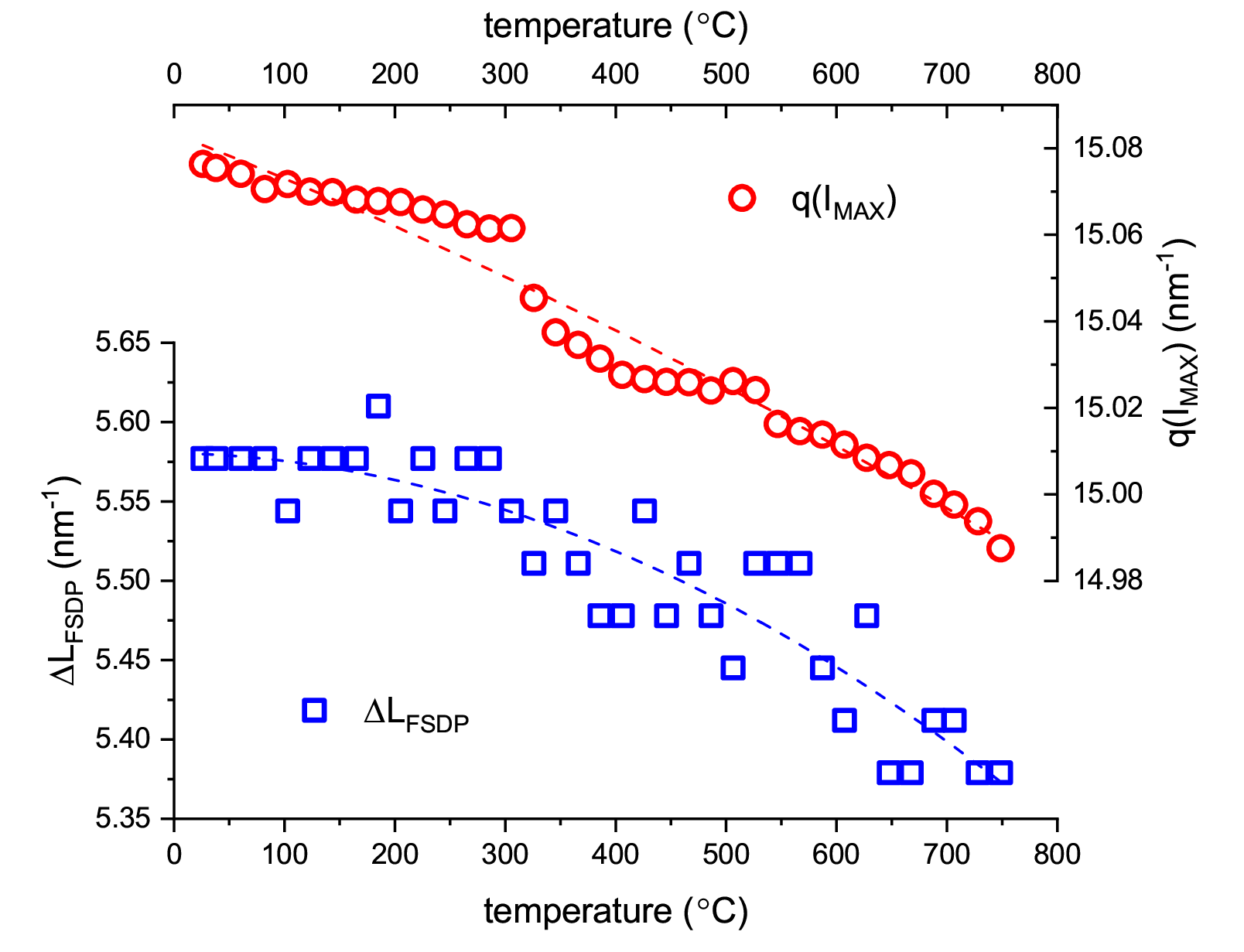}
\caption{Evolution of the width $\mathrm{\Delta L_{FSDP}}$ and the position of the maximum of the FSDP $\mathrm{q \left( I_{MAX} \right) }$ of pristine silica as a function of temperature. The dashed line are parabolic fit of the data, act as guide for the eyes and represent the evolution of the mean of the data.}
\label{fig:sio2_pristine}
\end{figure}

We reviewed the behavior of the FSDP of pristine silica with temperature. The FSDP of silica, recorded by both neutron and X-ray diffraction, is known not to evolve significantly with respect to the temperature within our temperature range.\cite{Susman1991} Quantitatively, Susman \textit{et. al.} observed a shift of $\mathrm{0.1 \ nm^{-1}}$ toward low q and and a decrease of $\mathrm{\Delta L_{FSDP}}$ of $\mathrm{0.1 \ nm^{-1}}$ when the temperature increases from ambient to 1036 \degree C.\cite{Susman1991} Our results for the evolution of the position of the maximum and the width of the first sharp diffraction as a function of temperature for pristine silica are shown in figure \ref{fig:sio2_pristine}.
The position of the FSDP at ambient temperature is $\mathrm{0.1 \ nm^{-1}}$ below literature data.\cite{Meade1992, Sato2008} Comparison of $\mathrm{\Delta L_{FSDP}}$ with other study is irrelevant because, to our knowledge, the way the HWHM is extracted is never clearly detailed.

When the temperature rises, we observe the same trends than Susman \textit{et. al.}, although very slightly more pronounced. Over an increase of temperature from ambient to 800 \degree C, so 200 \degree C below Susman \textit{et. al.} study, we see a shift toward low q of $\mathrm{0.1 \ nm^{-1}}$ and a decrease of $\mathrm{\Delta L_{FSDP}}$ of $\mathrm{0.2 \ nm^{-1}}$, both progressive. The examination of the data show a dispersion relatively to the mean (dashed lines in figure \ref{fig:sio2_pristine}). These fluctuations are either completely random for $\mathrm{\Delta L_{FSDP}}$, or more complex for $\mathrm{q \left( I_{MAX} \right) }$. Such fluctuations have multiple origins, from the raw data to the fitting process, so they are representative of the overall uncertainties on our results. Thus, for this study, we used the uncertainties derived here, i.e. $\mathrm{\pm 0.01 \ nm^{-1}}$ for $\mathrm{q \left( I_{MAX} \right) }$ and $\mathrm{\pm 0.05 \ nm^{-1}}$ for $\mathrm{\Delta L_{FSDP}}$.

\section{Density evolution\\}
\begin{figure}[h!]
\center
\includegraphics[width=8cm]{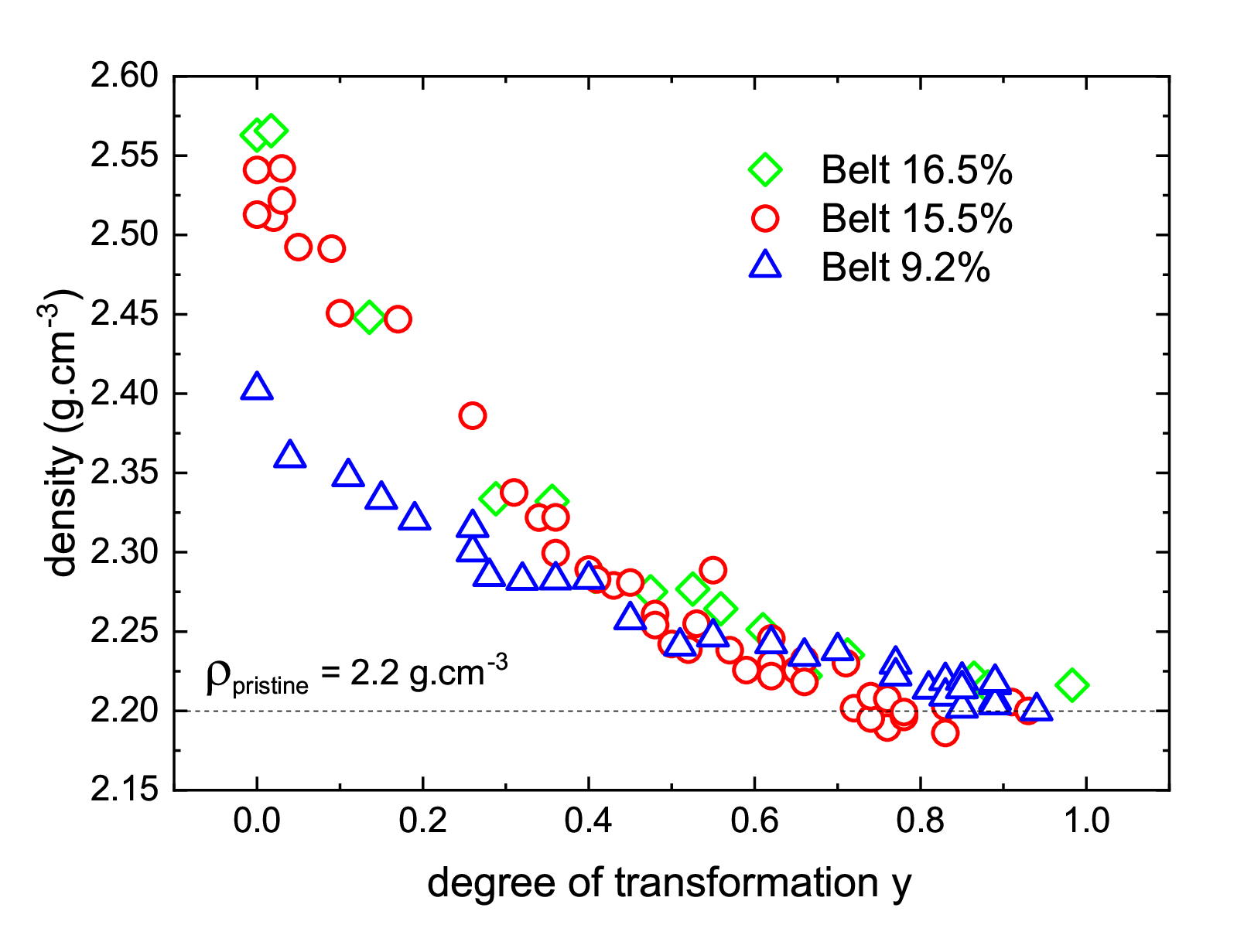}
\caption{Evolution of the density for the three different samples during the full relaxation process, expressed as a function of the degree of transformation.}
\label{fig:density}
\end{figure}

The evolution of the density for the three different densified samples during the complete relaxation is shown in figure \ref{fig:density}. The densities were extracted during previous Raman experiments\cite{Cornet2017}. Briefly, images of the full samples are recorded under a microscope during heating and isothermal annealing, the contour of the sample is later extracted and the surface computed. The surface data are then transformed into a volume assuming an isotropic expansion. The data are finally normalized to the density of the densified samples obtained prior heating by a buoyancy method.
The thermal expansion coefficient of silica being very low\cite{Bruckner1970} ($\mathrm{4.8 \leqslant ... \leqslant  5.1 \times 10^{-7} \ K^{-1}}$ over 0-900\degree C), the density changes visible in figure \ref{fig:density} are due to the relaxation processes only.

\section{Expression of Raman D2 data with respect to temperature\\}

\begin{figure}[h!]
\center
\includegraphics[width=9cm]{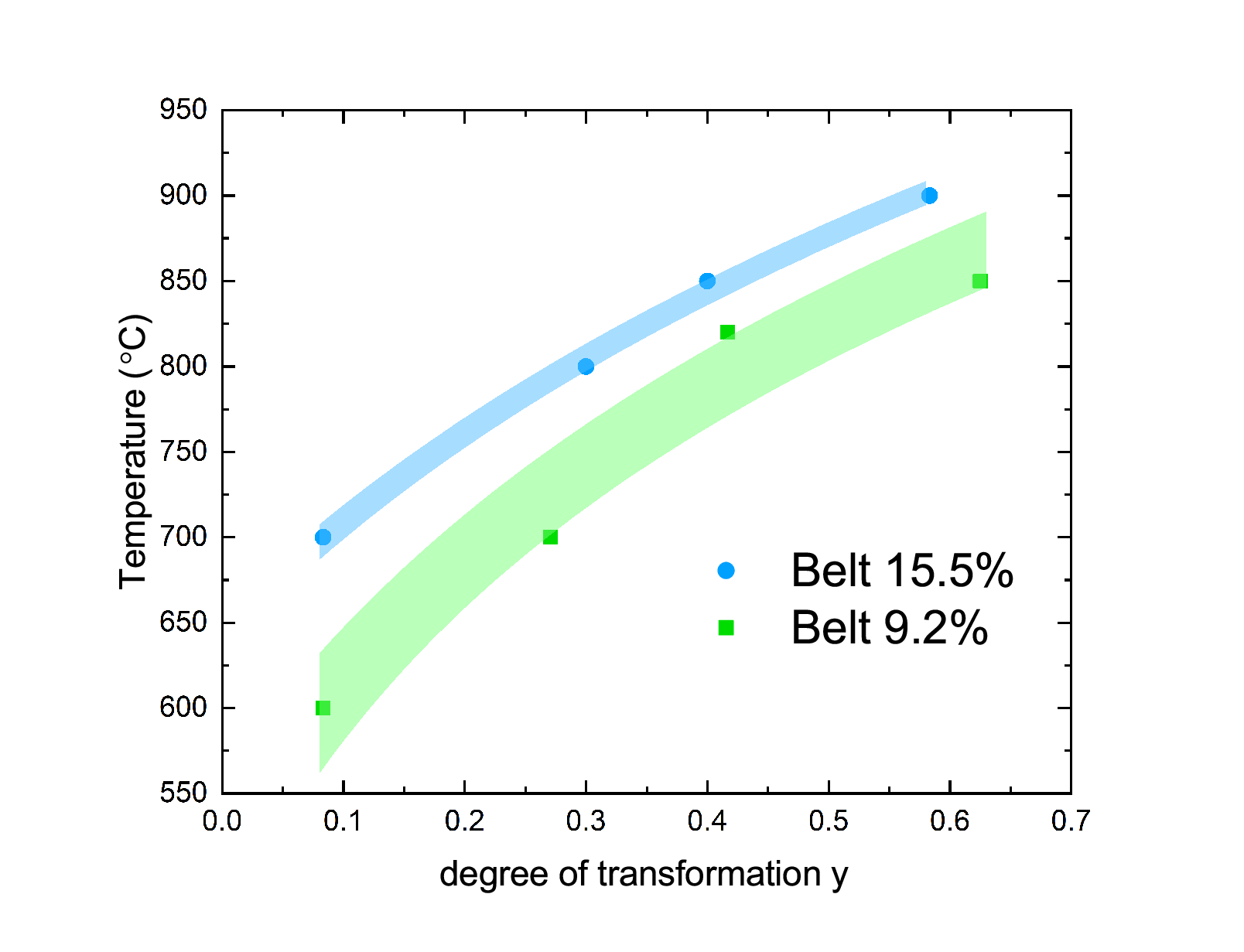}
\caption{Record of the temperature at which a certain degree of transformation is reached for the Raman experiments on Belt 9.2\% and Belt 15.5\% samples. The filled areas represent the standard deviation, obtained from a fit to an Arrhenius exponential law.}
\label{fig:y_T}
\end{figure}
In our previous paper, we monitored the isothermal annealings of the samples only, using Raman spectroscopy. To follow the entire reaction, we performed a complete set of experiments with annealings at temperature ranging from 500\degree C to 900\degree C. Finally, we were able to express our results, regarding the D2 band of the Raman spectrum, as a function of a degree of transformation $y$ ($0 \leqslant y \leqslant 1$). In the current study, we analyze the heating ramp as well. Moreover, we find that the state of maximum disorder, i.e. the signature of the polyamorphic transition, appears during that heating ramp for the samples Belt 9.2\% and Belt 15.5\%. In order to compare the Raman results to our present study, we need to express the D2 maximum as a function of temperature. Thus we need to express the temperature as a function of degree of transformation $y$. In our previous paper, as we performed only isothermal annealings, we had to consider the part of the relaxation happening during the heating ramp. We found that this lead to an uncertainty on the relaxation time that is negligible. Thus, we can use the first values of $y$ in each isothermal annealing to establish the relation between $y$ and the temperature. This is pictured in figure \ref{fig:y_T}, with the filled areas corresponding to the standard deviation.
\begin{figure}
\center
\includegraphics[width=9cm]{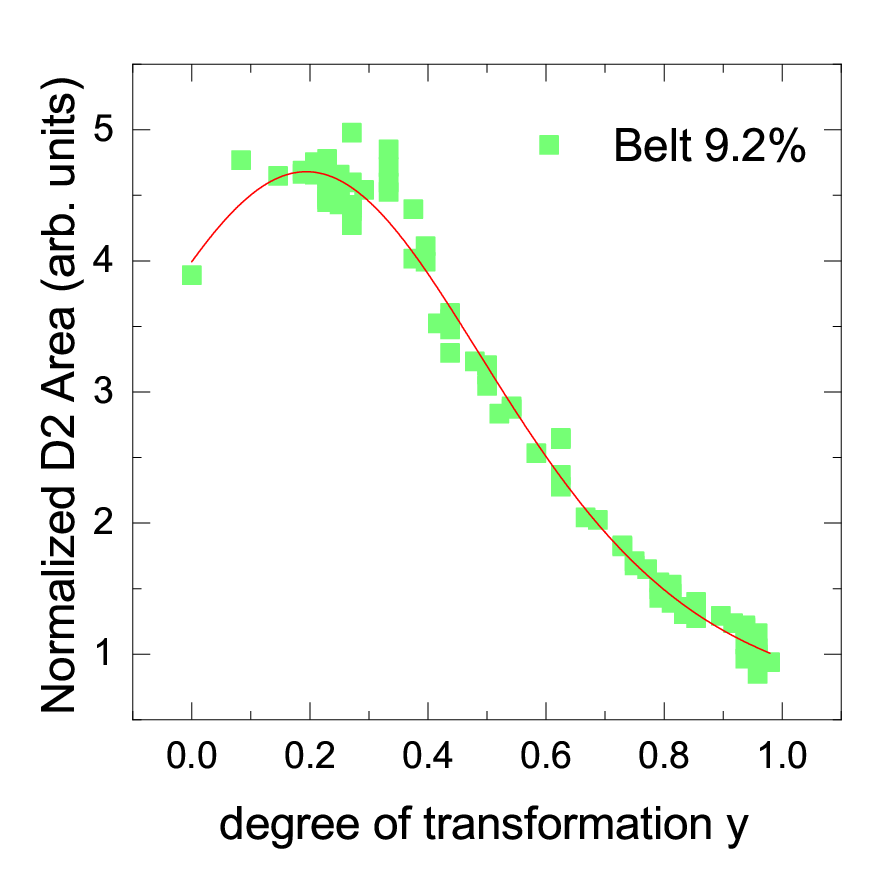}
\caption{Evolution of the Raman D2 band area as a function of the degree of transformation for the Belt 15.5\%, from ref \onlinecite{Cornet2017}.}
\label{fig:D2_9_2}
\end{figure}
\begin{figure}
\center
\includegraphics[width=9cm]{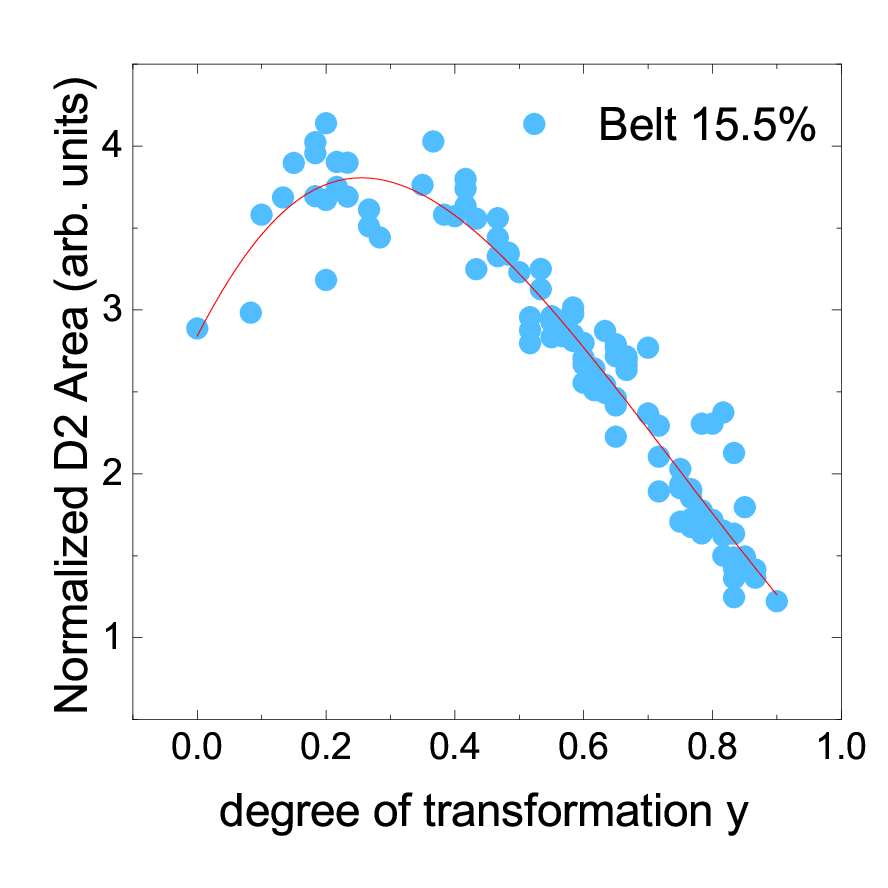}
\caption{Evolution of the Raman D2 band area as a function of the degree of transformation for the Belt 15.5\%, from ref \onlinecite{Cornet2017}.}
\label{fig:D2_15_5}
\end{figure}
In the figures \ref{fig:D2_9_2} and \ref{fig:D2_15_5} we represent the data of the normalized area of the Raman D2 band for the experiments led on Belt 9.2\% and Belt 15.5\% samples. For a complete interpretation of that result, refer to our previous paper\cite{Cornet2017}. However, the maximum in these data indicates, as for SAXS and WAXS data, the presence of an transitory state characterized by a higher disorder at the length scale corresponding to the D2 band. The intervals considered for the determination of the maximum of the D2 area are $\mathrm{y = 0.18-0.23}$ and $\mathrm{y = 0.22-0.29}$ for Belt 9.2\% and Belt 15.5\% respectively.

Using both figures, we can estimate the temperatures at which the maximum in the D2 band area appears for the two samples. We find $T = 688 \pm 40 \degree C$ for the Belt 9.2\% sample and $T = 780 \pm 40 \degree C$ for the Belt 15.5\% sample.

\bibliography{biblio}